\documentclass[copyright,creativecommons]{eptcs}
\providecommand{\event}{GandALF 2013} 
\usepackage{breakurl}             

\usepackage{graphicx}

\usepackage{hyperref}

\usepackage[caption=false]{subfig}
\usepackage{xspace}
\usepackage{amsmath}
\usepackage[noend]{algorithmic}
\usepackage{algorithm}
\usepackage{amssymb}
\usepackage{tikz}
\usetikzlibrary{arrows}
\usepackage{paralist}

\newcommand{\cost}{qty}
\newcommand{\N}{\mathcal{AG}}
\newcommand{\Nat}{{\mathbb N}}

\newcommand{\ltl}{\ensuremath{\mathsf{LTL}}\xspace}
\newcommand{\ctl}{\ensuremath{\mathsf{CTL}}\xspace}
\newcommand{\ctlstar}{\ensuremath{\ctl^*}\xspace}
\newcommand{\cl}{\ensuremath{\mathsf{CL}}\xspace}
\newcommand{\atl}{\ensuremath{\mathsf{ATL}}\xspace}
\newcommand{\atlstar}{\ensuremath{\atl^*}\xspace}
\newcommand{\atlplus}{\ensuremath{\atl^+}\xspace}
\newcommand{\rbcl}{\ensuremath{\mathsf{RBCL}}\xspace}
\newcommand{\rbatl}{\ensuremath{\mathsf{RB\text{-}ATL}}\xspace}
\newcommand{\prbatl}{\ensuremath{\mathsf{PRB\text{-}ATL}}\xspace}
\newcommand{\ral}{\ensuremath{\mathsf{RAL}}\xspace}
\newcommand{\ralstar}{\ensuremath{\ral^*}\xspace}

\newcommand{\team}[1]{\ensuremath{\langle\langle #1 \rangle\rangle}\xspace}
\newcommand{\market}[2]{\ensuremath{{#1 \vec{#2}}}\xspace}

\newcommand{\money}{\ensuremath{\$}}

\newcommand{\mossa}[1]{\ensuremath{[#1]}}

\newcommand{\lbatm}{\ensuremath{\mathsf{LB\text{-}ATM}}\xspace}

\newcommand{\atminputsym}{\ensuremath{\Sigma}\xspace}
\newcommand{\atmtapesym}{\ensuremath{\Gamma}\xspace}

\newcommand{\atmstate}[1]{\ensuremath{\mathsf{#1}}\xspace}
\newcommand{\atminitstate}{\ensuremath{\atmstate{q_0}}\xspace}
\newcommand{\atmstateset}{\ensuremath{\mathcal{Q}}\xspace}

\newcommand{\atmforallstateset}{\ensuremath{\atmstateset_{\forall}}\xspace}

\newcommand{\atmexiststateset}{\ensuremath{\atmstateset_{\exists}}\xspace}

\newcommand{\atmconfset}{\ensuremath{\mathcal C}\xspace}
\newcommand{\atmnextconf}[1]{\ensuremath{\atmconfset_{next(#1)}}\xspace}

\newcommand{\atmtapeconf}[1]{\ensuremath{\mathsf #1}\xspace}

\newcommand{\fullstate}[2]{\ensuremath{\langle #1, #2 \rangle}\xspace}

\newcommand{\execstep}[3]{\ensuremath{#1 \xrightarrow{#2} #3}\xspace}

\newcommand{\instrset}{\ensuremath{\mathcal{I}}\xspace}

\newcommand{\atmconf}[2]{\ensuremath{(#1, #2)}\xspace}
\newcommand{\atmaction}[3]{\ensuremath{\langle #1, #2, #3 \rangle}\xspace}
\newcommand{\atminstr}[2]{\ensuremath{#1 \rightarrow #2}\xspace}

\newcommand{\ignore}[1]{}

\newcommand{\ignoreconference}[1]{}

\newcommand{\mybox}[5]{

\draw #1 node(#2.west) {};

\draw #1 node(#2)[right,rectangle,minimum height=#5*5mm, minimum width=10mm, draw, rounded corners]{
#3
};

\draw node[above] at (#2.north) {
#4
};

\foreach \i in {1,...,#5} {
  \draw (#2.north east) ++(0,-\i) node[circle,draw,above=3mm,left](#2_out\i) {};
}
}

\newcommand{\myboxCustom}[6]{

\draw #1 node(#2.west) {};

\draw #1 node(#2)[right,rectangle,minimum height=#6, minimum width=10mm, draw, rounded corners]{
#3
};

\draw node[above=-1.3mm] at (#2.north) {
#4
};

\foreach \i in {1,...,#5} {
  \draw (#2.north east) ++(0,-\i/2.5) node[circle,draw,left](#2_out\i) {};
}
}

\newtheorem{definition}{Definition}
\newtheorem{theorem}{Theorem}
\newtheorem{example}{Example}
\newtheorem{corollary}{Corollary}

\title{Model checking coalitional games in shortage resource scenarios\footnote{The
work of Dario Della Monica
has been partially supported by the project \emph{Processes and Modal Logics}
(project nr.~100048021) of the Icelandic Research Fund and
the project \emph{Decidability and Expressiveness for Interval Temporal Logics}
(project nr.~130802-051) of the Icelandic Research Fund in partnership with the
European Commission Framework 7 Programme (People) under ``Marie Curie Actions''.
The work of Margherita Napoli
has been partially supported by the Italian PRIN 2010 project \emph{Logical Methods of Information Management}.
The work of Margherita Napoli and Mimmo Parente
has been partially supported by the Italian  FARB  projects 2010-2012.
}}
\author{Della Monica, Dario
\institute{
ICE-TCS, School of Computer Science\\
Reykjavik University, Iceland}
\email{dariodm@ru.is}
\and
Napoli, Margherita
\institute{
Dipartimento di Informatica\\
University of Salerno, Italy}
\email{napoli@dia.unisa.it}
\and
Parente, Mimmo
\institute{
Dipartimento di Informatica\\
University of Salerno, Italy}
\email{parente@unisa.it}
}
\def\titlerunning{Model checking coalitional games in shortage resource scenarios}
\def\authorrunning{D. Della Monica, M. Napoli, M. Parente}
\begin{document}
\maketitle

\begin{abstract}
Verification of multi-agents systems (MAS) has been recently studied taking into account the need of expressing resource bounds. 
Several  logics  for specifying properties of MAS    have been presented in quite a variety of scenarios with bounded resources. 
In this paper, we study a different formalism,  called \emph{Priced Resource-Bounded Alternating-time Temporal Logic}
(\prbatl),  whose  main novelty consists in moving  the notion of resources from a syntactic level (part of the formula) to a semantic one (part of the model). This allows us to track the evolution of the resource availability along the computations and provides us with a formalisms capable to model a number of real-world scenarios.  Two relevant aspects are the notion of global availability of the resources on the market, that are shared  by the agents,  and the notion of price of resources, depending on their availability.   In a previous work of ours, an initial step towards this new formalism was introduced, along with an EXPTIME algorithm for the model checking problem. In this paper we better analyze the features of the proposed formalism, also in comparison with previous approaches. The main technical contribution  is the proof of the EXPTIME-hardness of the the model checking problem for \prbatl, based on a reduction from the acceptance problem for \emph{Linearly-Bounded Alternating Turing Machines}. In particular, since the problem has multiple parameters, we show two {\em fixed-parameter} reductions.
\end{abstract}

\section{Introduction}
Verification of multi-agents systems (MAS) is a topic under investigation 
by several research groups in computer science in the last ten years
(\cite{Dastani:2010:SVM:1855030}).
Most of the research is based on logical formalisms, maybe
the most famous being the \emph{Alternating-time Temporal Logics} (\atl) \cite{AHK02} 
and the \emph{Coalition Logic} (\cl)
\cite{Pauly01,Pauly02}, both oriented towards the description of
collective behaviors
and used as specification languages for open systems.
These scenarios are hence naturally modeled as games. In \cite{Goranko01} it has
been shown that \cl can be embedded into \atl.
Recently, these two logics
have been used for the verification of multi-agent systems (MAS),
enhanced with resource constraints \cite{ALNR09,ALNR10,BNFN09,BF10,dnp11}.
The intuitive idea is that agent actions consume and/or produce
resources, thus the choice of a given action of an agent
is subject to the availability of the resources.
In \cite{ALNR09}, Alechina et al.
introduce the logic \emph{Resource-Bounded Coalition Logic} (\rbcl),
whose language extends the one of \cl
with explicit representation of resource bounds.
In \cite{ALNR10},
the same authors propose an analogous extension for \atl, called
\emph{Resource-Bounded Alternating-time Temporal Logics } (\rbatl),
and give a model checking procedure that runs in time $O(| \varphi
|^{2 \cdot r + 1} \times |G|)$, where $|\varphi|$ is the length of the formula $\varphi$ 
to be checked, $|G|$ is the size of the model $G$, and $r$ is the number of resources. 
However, the problem of determining a lower bound to the model checking problem 
is left open.
In \cite{BF10}, Bulling and Farwer introduce two \emph{Resource-Bounded Agent Logics}, called \ral and
\ralstar. The former represents a generalization of Alechina et
al.'s \rbatl, the latter is an analogous extension of \atlstar
(analogous extensions for, respectively, \ctl and \ctlstar were presented by
the same authors in \cite{BNFN09}).
The authors study several syntactic and semantic variants
of \ral and \ralstar
with respect to the (un)decidability of the model checking problem. 
In particular, while previous approaches
only conceive actions {\em consuming} resources, they introduce the
notion of actions {\em producing} resources. It turned out that such
a new notion makes the model checking problem undecidable.
Formulae of the
formalisms proposed in \cite{ALNR09,ALNR10,BNFN09,BF10} allow one to
assign an endowment of resources to the agents by means of the
so-called \emph{team operators} (borrowed from \atl). The problem is
then  to determine whether the agents in the {\em proponent} team have a strategy for the game to carry out
the assigned goals with that bounded amount of resources, whatever the agents in the
{\em opponent} team do.

In this paper we study a different formalism, called
\emph{Priced Resource-Bounded Alternating-time Temporal Logic} (\prbatl), introduced in
\cite{dnp11}, but in a much less mature version.
The key features of this new approach toward the formalization of such complex systems can be
summarized as follows.

\begin{compactitem}

 \item \emph{Boundedness of the resources.}
This is a crucial point in our formalization.
In order to model boundedness of the resources, a notion of \emph{global availability} of
resources on the market (or in nature), which evolves depending on
both proponent and opponent behaviors, is introduced.
Such a global availability is a semantic component (it is part of the structure where the logic
is interpreted) and its evolution is tracked during the executions of the system.
Agents' moves are affected by the current global availability
(e.g., agents cannot consume an unbounded amount of resources).

\smallskip

\item \emph{Resources are shared.}
Resources are global, that is, they are shared by all the agents. 
Thus, the  agents either consume or produce resources out of a shared pool of bounded capability,
and acquisition (resp., release) of a resource by an
agent (independently if the agent belongs to the proponent or
opponent team) implies that the resource will be
available in smaller (resp., greater) quantity.
In this way, we can model several scenarios where shared resources are acquired at a cost
that depends on that resource current availability
(for example in concurrent systems where there is a competition on resources).

\smallskip

\item \emph{Money as a meta-resource.}
%
In addition to public shared resources, our setting also allows one to model
\emph{private} resources, that is, resources that are possessed by agents
(public resources are present in the market and will be acquired by the agents in
case they need). The idea is to provide the agents with the unique private resource,
\emph{money}, that can be used to acquire (public) resources needed to perform the tasks.
In this sense, money represent several resource combinations and can be considered 
as a meta-resource.
Unlike the other resources, it is a syntactic component (money endowment is part of the formula),
and is the only (meta-)resource which is private for an agent.
 
At this stage, our formalization only features the possibility of assigning to agents
one private resource.
Nevertheless, in principle, it is possible to extend the idea to
admit a vector of private resources.
Furthermore, one could think of including the same resource in both the
pool of public resources and in the pool of private ones.
For instance, in a car race one of the players (the cars) possesses some gasoline in the tank
(private resource) but he needs to acquire more gasoline at the gas station (public resource)
to complete the race.

\smallskip

\item \emph{Resource production.}
Production of resources is allowed
in a quantity that is not greater than a fixed amount.
Thus, we  extend the model still preserving the decidability of the model checking problem. 
Observe that
the constraint  we impose  still allows us to describe many interesting real-world  
scenarios, such as acquiring memory by a program, or  leasing a car during a travel,
or, in general, any release of resources previously acquired.
A similar setting has been already observed also in \cite{BF10}.

\smallskip

\item \emph{Opponent power.}
First observe that we use the standard terminology which separates the role of the agents in a proponent 
team and those in the opponent team. This distinction is 
not within the game structure, but it is 
due to the formula   under consideration. 
%
Agents of the opponent team are subject to resource availability in choosing the action to perform,
in the same way as the agent of the proponent team, thus  the opponent team cannot interfere with a proponent strategy
performing  actions which either consume or produce too much (see Example \ref{ex:feasible_again} in Section  \ref{sec:preliminaries}).
However, it is common practice to consider opponent having maximum power,
to look for robust strategy.
We give  unlimited economic power to the agents in the opponent team, in the sense that
at each moment they have money enough 
to acquire the resources they need for a move, provided that the resources are available.

\end{compactitem}

Actually in \cite{dnp11} an EXPTIME algorithm for the model checking problem was given, 
along with a PSPACE lower bound.
%
%
The main technical contribution here is to provide an EXPTIME lower bound
for the model checking problem for \prbatl.
This result shows that the model checking problem 
for this logic is EXPTIME-complete.
The hardness proof is obtained by means of a reduction from
the acceptance problem for \emph{Linearly-Bounded Alternating Turing Machines} (\lbatm),
known to be EXPTIME-complete \cite{CKS81},
to the model checking problem for \prbatl.
More precisely, let $n$ be the number of agents, $r$ the number of resources,
and $M$ the maximum component occurring in the initial resource availability vector,
the algorithm given in \cite{dnp11} runs in exponential time in $n$, $r$, and
 the size of the representation of $M$ (assuming that $M$ is represented in binary).
To prove here the inherent difficulty with respect to multiple input parameters,
we show two reductions:
one parametric in the representation of $M$ (the digit size), that assumes constant both $n$
and $r$, and another parametric in $r$, and assuming constant both $n$ and the value of $M$.

%

\section{Comparison with related works}

In this section we compare our approach with the existing literature
underlining differences and similarities respect to  ~\cite{ALNR10} and \cite{BF10}.

%
%
In the work by Alechina et al.~\cite{ALNR10}, 
resource bounds only appear in the
formulae and are applied solely to the proponent team, but they are
not represented inside the model. 
Indeed, agents of the proponent team are endowed with new
resources at the different steps of the system execution.
This means that it is possible to ask whether a team can reach a goal with a given amount
of resources, but it is not possible to keep trace of the evolution of the global availability
of resources. Moreover,  resources are private to agents of the proponent team (not 
shared, as in our approach)  and  resource
consumption due to the actions of the opponent is not controlled.
%
Here instead, we keep trace of resource global availability, whose evolution depends
on both proponent and opponent moves.
%
In this way, it is possible to avoid undesired/unrealistic computations of the system
such as, for instance, computations consuming unboundedly.
%
Let us see a very simple example.
%
Consider the formula $\psi = \team{A^{\vec{\money}}} \Box p$. Its  semantics 
is that agents in team $A$ have together a strategy which can guarantee that $p$ always holds, 
whatever agents of the opponent team do (without consuming too many resources)
and provided the expense of the agents in $A$ does not exceed  $\vec{\money}$. 
A loop in the structure where the joint actions of agents consume resources without producing them, 
cannot be a model for $\psi$. 
On the contrary, consider the formula $\psi' = \team{A^{b}} \Box p$, belonging
to the formalism proposed in \cite{ALNR10}, expressing a similar property, with the only difference that 
the agents of $A$ use an amount of resources bounded by $b$.
A model for $\psi'$ must contain a loop where the  actions of agents in $A$
do not consume resources, but the  actions of agents in the opponent team
may possibly consume resources, leading to an unlimited consumption of resources.

%
%
%


As a further difference, recall that in   \cite{ALNR10} actions can only consume resources.
%
Without resource productions,  the model for many formulae 
(for example those containing the \emph{global}   operator $\Box$)
must have a loop  whose actions do not consume resources (\emph{do-nothing} actions), 
and a run satisfying these formulae  is eventually formed by only such actions. 
On the contrary, by allowing resource production, we can model more complex situations when   dealing with infinite games.


Finally, observe that a similarity with the cited paper  is  in the role of money, that could be seen as a private resource, endowed to the agents of the proponent team. 


Bulling and Farwer \cite{BF10} adopted an ``horizontal'' approach,
in the sense that they explored a large number of variants of a formalism to
model these complex systems.
In particular, they explored the border between decidability and undecidability of
the model checking problem for all such variants, and they showed how the status
of a formalisms (wrt decidability of its model checking problem) is affected by
(even small) changes in language, model, and semantics.
Our work takes advantage of this analysis in order to propose a logic that captures
several desirable properties (especially concerning the variety of natural real world scenario
that is possible to express), still
preserving decidability.
However,  our approach presents  conceptual novelties that
make it difficult to accomplish a direct comparisons between the formalisms
presented here and the ones proposed in \cite{BF10}.
We are referring here to both  the above mentioned idea of dealing with resources as
global entities for which agents compete, and  the notion of cost of
resource acquisition (price of the resources) that dynamically changes depending on
the global availability of that resource (thus allowing one to model the classic
market law that says that getting a resource is more expensive in shortage scenario). 
In \cite{BF10}, there is no such a notion as resources are assigned to (team of) agents
and proponent and opponent do not compete for their acquisition.
%

\medskip

As regards the complexity issue, in \cite{BF10}, no complexity analysis
(for the model checking problem) is performed,
while, in \cite{ALNR10}, an upper bound is given for \rbatl, that matches the
one given in \cite{dnp11} for \prbatl.  The algorithm  for \prbatl runs in exponential time
in  the number  $n$ of agents, the number $r$  of resources,
and the digit size of  the maximum component  $M$ occurring in the initial resource availability vector (assuming a binary reppresentation). 
Analogously, the model checking algorithm 
for \rbatl\ runs in  exponential time in $r$, in
 the digit size of 
 the maximum component of
resource endowment vectors  $b$ occuring in team  operators $\team{A^{b}} $ of  $\varphi$ 
and in the number $n$ of the agents (this is implicit  in set of states of $|G|$). 
Actually, both $n$ and $r$ are often treated as constant \cite{ALNR10,AHK02} 
(without this assumption, the complexity of \atl model-checking
is shown to be exponential in the number of agents \cite{DBLP:conf/ceemas/JamrogaD05}).
However, no complexity lower bound has been exhibit so far.
Aim of this paper is to fill this gap, by providing an
EXPTIME lower bound for \prbatl.


\section{A logical formalization: \texorpdfstring{\prbatl}{PRB-ATL}} \label{sec:preliminaries}

\noindent{\bf Syntax.}
We start with the introduction of some notations we will use in the
rest of the paper. The set of  \emph{agents}  is $\N = \{ a_1, a_2,
\ldots, a_n \}$ and a  \emph{team} is any subset of $\N$.
The integers  $n$ and $r$ will be used throughout the paper to denote the number of  agents
and \emph{resource types} (or simply \emph{resources}), respectively.
Let $\mathcal M = (\mathbb N \cup \{ \infty \})^{r}$
denote the set of
\emph{global availabilities of resources on the market (or in nature)}
and let $\mathcal N = (\mathbb N \cup \{ \infty \})^{n}$ denote the set of
\emph{money availabilities for the agents},
where  $\mathbb N$ is the set of natural numbers (zero included).
Given a money availability $\vec{\money} \in \mathcal N $, its $i$-th component
$\vec{\money}[i]$ is the money availability of agent $a_i$\footnote{Throughout all the
paper, symbols identifying vectors are denoted with an arrow on the top (e.g., $\vec{\money}$, $\vec{m}$).}.
Finally, the set $\Pi$ is a finite set of \emph{atomic propositions}.

The formulae of \prbatl are given by the following grammar:
$$ \varphi ::= p \mid \neg \varphi \mid \varphi \wedge \varphi \mid
\team{A^{\vec{\money}}} \bigcirc \varphi \mid \team{A^{\vec{\money}}} \varphi \mathcal U \varphi \mid
\team{A^{\vec{\money}}} \Box \varphi \mid \market{\sim}{m}
$$
where $p \in \Pi$, $A \subseteq \N$, $\sim \in \{<, \leq, =, \geq, >
\}$, $\vec{m} \in \mathcal M$ and $\vec{\money} \in \mathcal N$.
Formulae of the kind $\market{\sim}{m}$ test the
current availability of resources on the market.
As usual, other standard operators can be considered as abbreviation, e.g.,
the operator $\team{A^{\vec{\money}}} \Diamond \psi$ can be defined as
$\team{A^{\vec{\money}}} \top \mathcal U \psi$, for every formula $\psi$.

%
%

\medskip

\noindent{\bf Priced game structure.}
%
%
%
\emph{Priced game structures} are defined by extending the definitions
of concurrent game structure
and resource-bounded concurrent game structure given in, respectively, \cite{AHK02} and \cite{ALNR10}.

\begin{definition}
A \emph{priced game structure} $G$ is 
a tuple $ \langle  Q,  \pi, d, D, \cost, \delta, \rho, \vec{m_0} \rangle$, where:
\begin{compactitem}
\item
$Q$ is the finite set of \emph{locations}; $q_0 \in Q$ is called \emph{initial location}.
\item
$\pi: Q \rightarrow 2^{\Pi}$ is the \emph{evaluation function}, which determines the
atomic propositions holding true in each location.
\item
$d: Q \times \N \rightarrow \Nat$ is the \emph{action function}
 giving the number $d(q,a) \geq 1$ of \emph{actions} available to an agent $a \in \N$ at a location $q \in Q$.
The actions  available to $a$ at $q $ are identified with the numbers\footnote{No ambiguity will
arise from the fact that actions of different agents are identified with the same numbers.}
$1, \ldots, d(q,a)$ and a generic action is usually denoted by $\alpha$.
We assume that each agent has at least one available action at each location,
that could be thought of as the action \emph{do-nothing} and we assume that it is always the first.
\item
$D: Q \rightarrow 2^{\mathbb N^n} $
is a function that maps each location $q$ to the set of  vectors  $ \{ 1, \ldots, d(q,a_1) \} \times \ldots \times \{ 1, \ldots, d(q,a_n) \}$.
Each vector, called  \emph{action profile} and denoted by $\vec{\alpha}$, identifies a choice among 
the actions  available for each agent in the location $q$.
(The action of the agent $a$ in   $\vec{\alpha}$ is $\vec{\alpha}(a)$.)
\item
$\cost: Q \times \N \times \Nat \rightarrow \mathbb Z^r$ is a partial function,
where $\cost(q, a, \alpha)$, with $1\leq \alpha \leq d(q,a)$,
defines at location $q$ the amount of resources required by the $a$'s action $\alpha$.
We define $\cost(q,a,1) = \vec{0}$, that is the vector whose components are all equal to
$0$, for every $q \in Q$, $a \in \N$ (doing nothing neither consumes nor produces resources).
\item
$\delta: Q \times \Nat^n \rightarrow Q$ is the \emph{transition function}. For $q \in Q$ and
 $\vec{\alpha} \in D(q)$,
$\delta(q,\vec{\alpha})$ defines the next location reached from $q$ if the agents perform the
actions in  the action profile $\vec{\alpha} $.
\item
$\rho: \mathcal M \times Q \times \N \rightarrow \mathbb N^r$ is the
\emph{price function}. It returns the \emph{price vector}
of the resources (a price for each resource), based on the current resource availability
and location, and on the acting  agent.
\item
$\vec{m_0} \in \mathcal M$ is the initial global availability of resources.
It represents the resource availability on the market at the initial state of the system.
\end{compactitem}
\end{definition}

Note that a negative value in $\cost(q, a, \vec{\alpha})$
represents a resource consumption,
while a positive one represents a resource production.
We  also consider the extension of the function $\cost$, called again with the same name, to get the amount
of resources required by a given team. Thus,
for a location $q$, a team $A$ and an  action profile $\vec{\alpha} $,
 $\cost(q, A,  \vec{\alpha}) = \sum_{a \in A}\cost(q, a, \vec{\alpha}(a))$.
Moreover,
%
we will use the function $consd: Q \times \N \times \Nat \rightarrow \Nat^r$ that
for the tuple $(q, a, \alpha)$ returns the vector of the  resources which are consumed by an agent $a$,
being in state $q$, for an action $\alpha$. This vector is obtained from
$\cost(q, a, \alpha)$ by replacing the positive components, representing
a resource production,  with zeros,
and the negative components, representing a resource consumption, with their absolute values.

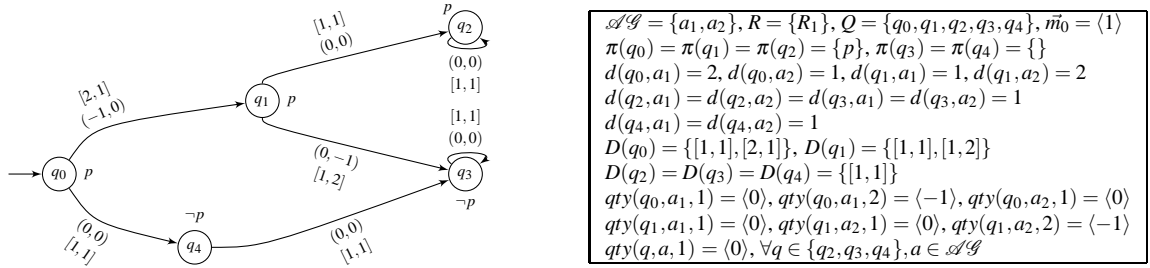
\begin{figure}
 \begin{tikzpicture}
\tiny
\begin{scope}[yscale=.65,xscale=.9]
\draw (-1,1) ++(9.2,0);

  \draw(0,0) node(q0)[circle,draw]{$q_0$};
  \draw (q0) node[right=2.5mm] {$p$};
  \draw[-latex'] (-.75,0) -- (q0);
  \draw(3,1.5) node(q1)[circle,draw]{$q_1$};
  \draw (q1) node[right=2.5mm] {$p$};
  \draw(2,-1.5) node(q4)[circle,draw]{$q_4$};
  \draw (q4) node[above=2.5mm]{$\neg p$};
  \draw[-latex'] (q0) .. controls++(0.5,1) .. (q1) node[pos=.5,sloped,above=.3cm]{$\mossa{2,1}$} node[pos=.5,sloped,above=.2]{$(-1, 0)$};
  \draw[-latex'] (q0) .. controls++(0.5,-1) .. (q4) node[pos=.5,sloped,below=.3cm]{$\mossa{1,1}$} node[pos=.5,sloped,below=.2]{$(0, 0)$};
  \draw(q1)++(3,1.5) node(q2)[circle,draw]{$q_2$};
  \draw (q2) node[above left=1mm] {$p$};
  \draw(6,0) node(q3)[circle,draw]{$q_3$};
  \draw (q3) node[below=2.5mm] {$\neg p$};
  \draw[-latex'] (q3) .. controls++(-.5,.5) and ++(.5,.5) .. (q3) node[pos=.5,sloped,above=.3cm]{$\mossa{1,1}$} node[pos=.5,sloped,above=.2]{$(0, 0)$};
  \draw[-latex'] (q1) .. controls++(0,.5) .. (q2) node[pos=.75,sloped,above=.3cm]{$\mossa{1,1}$} node[pos=.75,sloped,above=.2]{$(0, 0)$};
  \draw[-latex'] (q1) .. controls++(0,-0.5) .. (q3) node[pos=.75,sloped,below=.3cm]{$\mossa{1,2}$} node[pos=.75,sloped,below=.2]{$(0, -1)$};
  \draw[-latex'] (q2) .. controls++(-.5,-.5) and ++(.5,-.5) .. (q2) node[pos=.5,sloped,below=.3cm]{$\mossa{1,1}$} node[pos=.5,sloped,below=.2]{$(0, 0)$};
  \draw[-latex'] (q4) .. controls++(1,0) .. (q3) node[pos=.75,sloped,below=.3cm]{$\mossa{1,1}$} node[pos=.75,sloped,below=.2]{$(0, 0)$};
 
\end{scope}

\draw (10.75,.5)node{
\scriptsize
\begin{tabular}{|l|}
\hline
$\N = \{a_1, a_2\}$, $R = \{R_1\}$, $Q = \{q_0, q_1, q_2, q_3, q_4\}$, $\vec{m_0} = \langle 1 \rangle$ \\
$\pi(q_0) = \pi(q_1) = \pi(q_2) = \{p\}$, $\pi(q_3) = \pi(q_4) = \{ \}$ \\
$d(q_0,a_1) = 2$, $d(q_0,a_2) = 1$, $d(q_1,a_1) = 1$, $d(q_1,a_2) = 2$ \\
$d(q_2,a_1) = d(q_2,a_2) = d(q_3,a_1) = d(q_3,a_2) = 1$ \\
$d(q_4,a_1) = d(q_4,a_2) = 1$ \\
$D(q_0) = \{ \mossa{1,1}, \mossa{2,1} \}$, $D(q_1) = \{ \mossa{1,1}, \mossa{1,2} \}$ \\
$D(q_2) = D(q_3) = D(q_4) = \{ \mossa{1,1} \}$ \\
$\cost(q_0, a_1, 1) = \langle 0 \rangle $, $\cost(q_0, a_1, 2) = \langle -1 \rangle $, $\cost(q_0, a_2, 1) = \langle 0 \rangle $ \\
$\cost(q_1, a_1, 1) = \langle 0 \rangle $, $\cost(q_1, a_2, 1) = \langle 0 \rangle $, $\cost(q_1, a_2, 2) = \langle -1 \rangle $ \\
$\cost(q, a, 1) = \langle 0 \rangle$, $\forall q \in \{q_2,q_3,q_4\}, a \in \N$ \\
\hline
\end{tabular}
};
 \end{tikzpicture}

\caption{Example of priced game structure $G = \langle  Q,  \pi, d, D, \cost, \delta, \rho, \vec{m_0} \rangle$.}
\label{fig_game}
\end{figure}

\begin{example}\label{gamestructure}
A priced game structure with two agents $a_1$ and $a_2$ and one resource $R_1$ is depicted in \figurename~\ref{fig_game}.
The only atomic proposition is $p$, labeling the locations $q_0$, $q_1$, $q_2$.
The action profiles, labeling the transitions in the graph and depicted with square brackets, are as follows.
$D(q_0)= \{\mossa{1,1}, \mossa{2,1}\}$ is due to the existence of two actions of $a_1$
and one action of $a_2$
at location $q_0$, $D(q_1)= \{\mossa{1,1}, \mossa{1,2}\}$ corresponds to a single action of $a_1$
and two actions of $a_2$ at location $q_1$. In all the other locations the only action profile
is $\mossa{1,1}$ corresponding to the existence of a single action of both the agents.
The function $\cost$ is represented by parentheses.
The price vector is not depicted.
\end{example}

\noindent{\bf Semantics.}
In the following, given a resource availability $\vec{m}$, by $\mathcal M^{\leq \vec{m}}$ we denote the
set $\{ \vec{m'} \in \mathcal M \mid \vec{m'} \leq \vec{m} \}$.
In order to give the formal semantics let us first define the following notions.

\begin{definition}
A \emph{configuration} $c$ of a priced game graph $G$ is a pair $\langle q,
\vec{m} \rangle \in Q \times \mathcal M^{\leq \vec{m_0}}$.
Given two configurations
$c=\langle q, \vec{m} \rangle$ and $c'= \langle q', \vec{m'} \rangle$,
and an action profile $\vec{\alpha} \in D(q)$, we say that
$ c \rightarrow_{\vec{\alpha}} c'$ if
$q' = \delta(q, \vec{\alpha})$ and
 $ \vec{m'} = \vec{m} + \cost(q, \N, \vec{\alpha})$.
A \emph{computation}  over $G$ is an infinite
 sequence $C =  c_1 c_2 \ldots$ of configurations of $G$,
 such that
for each $i$
there is  an action profile  $\vec{\alpha}_i$ such that
$ c_i \rightarrow_{\vec{\alpha}_i} c_{i+1}$.
\end{definition}

Let $C =  c_1 c_2 \ldots$ be a computation. We denote  by $C[i]$
the $i$-th configuration $c_i$ in $C$  and by $C[i,j]$, with $1 \leq i \leq j $, the finite sequence of
configurations $c_i c_{i+1} \ldots c_{j}$ in $C$.
Given a configuration $c=\langle q,\vec{m} \rangle$ and a team $A$, a function  $\vec{\alpha_A}: A \rightarrow \Nat $
 is called   \emph{A-feasible in c} if
 there exists an action  profile $\vec{\alpha} \in D(q)$ with
$ \vec{\alpha_A}(a)= \vec{\alpha}(a)$ for all $a \in A$ and
$\vec 0 \leq \cost(q, A, \vec{\alpha}) + \vec{m} \leq \vec{m_0}$. In this case we say that 
$\vec{\alpha}$ \emph{extends} $\vec{\alpha_A}$.

\begin{definition}
A  \emph{strategy} $F_A$ of a team $A$  is a function which associates
to each finite sequence of configurations  $  c_1 c_2 \ldots c_s$,
a function  $\vec{\alpha_A}: A \rightarrow \Nat $ which is $A$-feasible in $c_s$.
\end{definition}

In other words, a strategy $F_A$ returns a choice of the actions of the agents in the team $A$,
 considering only those actions
whose resource consumption does not exceed the available amount and whose resource production
does not exceed the amount consumed so far.
Clearly, this constraint will limit both proponent and opponent team.

For  each  strategy $F_A$ of a  team $A$ and for each sequence of configurations $c_1 c_2 \ldots c_s$,
there are several possibilities for the next configuration $c_{s+1}$,
depending on the different  choices of the opponent team $\overline A = \N \setminus A$.
Anyway, fixed a strategy $F_{\overline A}$ of the opponent team, there is 
at most one action profile  obtained according to both the strategies, that is the action profile  $\vec{\alpha}$ extending both  $\vec{\alpha_A}$,   given by the strategy $F_A$, and
$\vec{\alpha_{\overline A}}$,   given by the strategy  $F_{\overline A}$ (i.e.
 $\vec{\alpha}$ is such that
 $\vec{\alpha}(a)=\vec{\alpha_X}(a)$, for $X \in \{A, \overline A\}$ and $a \in X$).
A computation $C = c_1,c_2 \ldots$,  is the  \emph{outcome of the strategies $F_A$ and
 $F_{\overline A}$  from the configuration $c_1$}  if, for each  $i \geq 1 $,
there is an action profile  $\vec{\alpha}_i$ obtained according to both $F_A$ and
 $F_{\overline A}$, such that
 $ c_{i} \rightarrow_{\vec{\alpha}_i} c_{i+1}$.
Given a strategy $F_A$ and a configuration $c$,  $out(c,F_A)$ denotes  the set of  the outcomes of $F_A$ and
 $F_{\overline A}$  from $c$, for all the strategies  $F_{\overline A}$ of the team  $\overline A$.
Observe that,  given a finite sequence of configurations $C = c_1 c_2 \ldots c_s$, if the action profile $ \vec{\alpha} $ according to the two strategies is
not such that $\vec{0} \leq \cost(q_s, \N, \vec{\alpha})  + \vec{m_s} \leq  \vec{m_0}$,
then there is no next configuration.
Thus  outcome of the strategies $F_A$ and  $F_{\overline A}$  from a given configuration
may be undefined (recall that we consider only infinite computations).

\begin{example}\label{unfeasible}
Consider the priced game structure in \figurename~\ref{fig_game}, with teams $A=\{a_1\}$ and $B=\{a_2\}$,
one resource type and initial global availability $\vec{m_0}=\langle 1 \rangle$.
Let $c=\langle q_0,\langle1\rangle \rangle$ be a sequence of configurations (of length $1$).
Team $A$ has two possible strategies in $c$, one for each possible action of agent $a_1$, and
team $B$ has one strategy for the single available action of agent $a_2$.
Suppose that, according to the strategy $F_A$, agent $a_1$ chooses to perform the action $2$ ($F_A(c)(a_1) = 2$), 
then the action profile \mossa{2,1} is performed and one unit of the unique resource is consumed.
In the obtained configuration $ \langle q_1,\langle 0\rangle \rangle$ the agent $a_1$ has one available action
while the agent $a_2$ has two actions.
Anyway $F_B$ cannot return the action $2$ for the agent $a_2$, since this action would require an amount of the
resource greater than $0$, which is the current availability.
Thus only the configuration  $\langle q_2, \langle0 \rangle \rangle$  can be reached and
the computation $C=\langle q_0,\langle1 \rangle \rangle \langle q_1, \langle 0\rangle \rangle \langle q_2, \langle 0\rangle \rangle \langle q_2, \langle0\rangle \rangle \ldots$ is the only one that belongs to $out(c,F_A)$.
\end{example}

%

Now we introduce the concept of consistent strategy. Two properties have to be satisfied:
first, the outcomes starting from $c$ are always defined and also the  agents of the proponent team have enough money to realize the chosen actions.


\begin{definition}
\label{def:strat}
Let $\vec{\money} \in \mathcal N$, $c$ be a configuration, $A \subseteq \N$ be the proponent team, and
$\overline A = \N \setminus A$ be the opponent team.
A strategy $F_A$ of $A$ is said to be \emph{consistent with respect to $\vec{\money}$ and $c$}
(\emph{$(\vec{\money} ,c)$-strategy}),  if
\begin{compactenum}
\item \label{cond1}
 for any strategy $F_{\overline A}$ of  $\overline A$, the
outcome of  $F_A$ and
 $F_{\overline A}$  from the configuration $c$ is defined,
\item \label{cond2}
for every  $C = c_1 c_2 \ldots \in out(c,F_A)$, with $c_i = \langle q_i, \vec{m_i} \rangle$,
for every $i \geq 1$ and $a_k \in A$:
$\sum_{j=1}^i \rho(\vec{m_j}, q_j, a_k) \cdot consd(q_j, a_k, F_A(C[1,j])(a_k)) \leq \vec{\money}[k]$.

\end{compactenum}
\end{definition}

In the above condition the dot operator denotes
the usual scalar product of vectors.
Observe that only the money availability of
the team $A$ is tested. Actually, we suppose that the opponent team $\overline A$ always has money enough to
make its choice.
Notice also that
the  actions {\em producing} resources do not cause a reimbursement
of money to the agents.
As it is usual when dealing with temporal logics,
we guarantee that priced game structures are non-blocking, in the sense that
at least a  $(\vec{\money} ,c)$-strategy  exists for a  given team $A$. Indeed, agents of  $A$ can always jointly choose the \emph{do-nothing} action.

A formula of \prbatl is evaluated with respect to a priced game structure
$G$ and a configuration $c = \langle q,\vec{m} \rangle$.
The definition of the semantics is completed by
the definition of the satisfaction relation $\models$:

%
%

\begin{compactitem}
 \item $(G, c) \models p$ iff $p \in \pi(q)$
 \item $(G, c)\models \neg \psi$ iff $ (G, c) \not\models \psi$
 \item $(G, c) \models \psi_1 \wedge \psi_2$ iff $(G, c) \models \psi_1$
 and $(G, c) \models \psi_2$
 \item $(G, c) \models \team{A^{\vec{\money}}} \bigcirc \psi$ iff there exists a
$(\vec{\money},c)$-strategy $F_A$ such that, for all $C \in out(c, F_A)$,
it holds that $(G,C[2]) \models \psi$
 \item $(G, c) \models\team{A^{\vec{\money}}} \psi_1 \mathcal U \psi_2$ iff there exists a
$(\vec{\money},c)$-strategy $F_A$ such that, for all $C \in out(c, F_A)$,
there exists $i \geq 0$ such that $(G,C[i]) \models \psi_2$
and, for all $1 \leq j < i$, it holds that $(G,C[j]) \models \psi_1$
 \item $(G, c) \models \team{A^{\vec{\money}}} \Box \psi$ iff there exists a
$(\vec{\money},c)$-strategy $F_A$ such that, for all $C \in out(c, F_A)$,
it holds that $(G,C[i]) \models \psi$ for all $i \geq  1$
 \item $(G, c) \models \market{\sim}{m'}$ iff $\vec{m} \market{\sim}{m'}$
where $\sim \in \{<, \leq, =, \geq, > \}$.
\end{compactitem}

\medskip
Given a \prbatl formula and a priced game srtucture $G$,
we  say that  $G$ satisfies $\varphi$, $ G\models \varphi$,  if
 $(G, c_0) \models \varphi$  where  $c_0 = \langle q_0, \vec{m_0} \rangle$.
The \emph{model checking} problem for \prbatl consists in verifying whether  $G \models \varphi$.

\begin{example}\label{ex:feasible_again}
Consider the priced game structure in \figurename~\ref{fig_game}, with teams $A=\{a_1\}$ and $B=\{a_2\}$.
A formula $\psi=\team{\N^{\vec{\money}}}\bigcirc\team{A^{\vec{\money'}}} \Box p$
holds true in the configuration
$\langle q_0, \langle 1 \rangle \rangle$, provided that  $\vec{\money}$ and   $\vec{\money'}$ are enough to make the move.
Indeed, $a_1$ and $a_2$ together are able to force the computation to reach the
$\langle q_1, \langle 0  \rangle \rangle$ (one unit of resource is consumed).
From such a configuration, the opponent team $B$ cannot force the computation into $q_3$, as
the action $2$ is not allowed for $a_2$ (no resources are available
to perform the action), and thus  $\psi$ holds.
Instead, $\psi$ is false in the configuration  $\langle q_0, \langle 2  \rangle \rangle$
(actually in each configuration $\langle q_0,  \langle x  \rangle \rangle$, with $x>1$),
because $\langle q_1, \langle 1 \rangle \rangle$ is reached after the execution of the first transition,
and in that configuration action $2$ for $a_2$ in $B$ is allowed, leading to
$q_3$.
Finally, notice that the formula is false also when evaluated in 
$\langle q_0, \langle 0 \rangle \rangle$, as the only possible transition is the one leading from $q_0$
to $q_4$ (no resources are available to perform action $1$ for agent $a_1$).
\end{example}

\section{Complexity lower bounds for the model checking problem}\label{hardnessMC}

In \cite{dnp11}, the authors presented an algorithm for model checking \prbatl,
providing an exponential upper bound for the problem.
In particular, let $n$ be the number of agents, $r$ the number of resources,
and $M$ the maximum component occurring in the initial resource availability vector,
the proposed algorithm runs in exponential time in $n$, $r$, and
the size of the representation of $M$
(assuming that $M$ is represented in binary).
In this section we prove that an algorithm
that behaves asymptotically better cannot exist, thus proving that the
problem is EXPTIME-complete.
To prove the inherent difficulty with respect to the multiple input parameters,
we show two reductions:
one parametric in the representation of $M$ (the digit size), which assumes both $n$
and $r$ constant, and the other parametric in $r$, this time
assuming constant both $n$ and the value of $M$.
We conjecture the existence of a third EXPTIME reduction, in which $r$ and $M$ are constant
and the parameter is $n$.
In fact, if it was not the case,
it would be possible to improve the proposed model checking
algorithm in a way that its complexity would not
be exponential in $n$.

We first recall the formalism of \emph{linearly-bounded
alternating Turing machines} (\lbatm) 
and the notion of \emph{hierarchical representation},
a succinct way of representing priced game structures
inspired to the work done in \cite{AY01} for classical Kripke structures.
Finally, we present the two reductions from the acceptance problem for \lbatm,
known to be EXPTIME-complete \cite{CKS81
}, to the model checking problem for \prbatl.

\subsection{Linearly-bounded alternating Turing Machines}\label{constAgAndRes}

A \emph{linearly-bounded alternating Turing machines} 
(\lbatm) is  a tuple $\langle \atmstateset, 
\atmtapesym,\instrset, \atminitstate, 
 \rangle$,
where \atmstateset is the set of \emph{states}, partitioned in 
$\atmforallstateset $ (\emph{universal states}) and  $\atmexiststateset $  (\emph{existential states});
$\atmtapesym$ is the set of \emph{tape symbols},
including the `blank' symbol $\mathsf B$, and two special symbols $\llcorner$ and $\lrcorner$, 
denoting the  left and right \emph{tape delimiters};
$\instrset \subseteq \atmstateset \times \atmtapesym \times
\atmstateset \times \atmtapesym \times \{ \leftarrow, \rightarrow\nobreak \}$ is the \emph{instruction set}; 
$\atminitstate \in \atmstateset$
is the \emph{initial state}.

 Symbols from $\atmtapesym$ are stored in the \emph{tape cells}, and  the first and the last cell of the tape store, respectively, the symbols $\llcorner$ and $\lrcorner$.
A \emph{tape configuration}  $\atmtapeconf{s}$ 
 is a sequence of the symbols  stored in the tape cells, and keeps trace of an \emph{head cell}. 
A \emph{configuration} $c$ is a pair  $ \atmconf{\atmstate q}{\atmtapeconf s}$
of a
state $ {\atmstate q}$ and a tape configuration ${\atmtapeconf s}$, and 
\atmconfset is the set of the configurations.
The initial configuration 
is  $c = \atmconf{\atminitstate}{\atmtapeconf{s_0}}$,
where  \atmtapeconf{s_0} contains the input, possibly  followed by a sequence of  blanks, and
its  head cell stores the first input symbol.

An \emph{instruction} $i = (\atmstate{q}, \lambda, \atmstate{r}, \nu ,\sim) \in \instrset $ is also denoted
\atminstr{\fullstate{\atmstate q}{\lambda}}{\atmaction{\atmstate r}{\nu}{\sim}},
where \fullstate{\atmstate q}{\lambda} is called a \emph{full state}.
Its  intuitive meaning 
is as follows: ``whenever the machine is in the state \atmstate q
and the symbol  in the head cell is $\lambda$, then the machine  switches
 to state \atmstate r, the symbol  in the head cell is replaced with 
$\nu$, and the head position is moved to the
left or to the right (according to $\sim$)". 
\ignoreconference{;
precisely, the tape length is equal to $f(n)$,
where $f$ is a linear function and $n$ is the length of the input.
This means that, given an input whose length is $n$, and 
the initial configuration 
$\atmconf{\atminitstate}{\atmtapeconf{s_0}}$, 
we have that $| \atmtapeconf{s_0} | = f(n)$.
To keep the length of the tape configuration constant during the
computation, instructions are such that whenever the symbol $\llcorner$
(resp., $\lrcorner$) is  in the head cell,
the same symbol is re-written, the machine switch to
a (possibly) new internal state, and the head moves to right (resp., to left).
Further, this is the only case a machine instruction writes
the symbol $\llcorner$ (resp., $\lrcorner$).
}
An \emph{execution step} of the machine is denoted
\execstep{c}{i}{c'}, where $c,c' \in \atmconfset$,
$i \in \instrset$ and $c'$ is the configuration
reached from $c$ after the execution of the instruction $i$.
Let $\atmnextconf{c}=
 \{c' \in \atmconfset \mid \execstep{c}{i}{c'}$ is an execution step,
for some $i \in \instrset \}$.
All the tape configurations are linear in the length of the input
and we follow the common practice to only consider machines whose tape length does not vary
 during the computation.
We can also assume that \lbatm have
no infinite computations since any \lbatm can be transformed into
another, accepting the same language and halting
in
a finite number of  steps.
Such a \lbatm counts
the number of execution steps and rejects any computation whose number of
steps exceeds the number of possible configurations.

The \emph{acceptance condition} is defined recursively.
A configuration $c = \atmconf{\atmstate q}{\mathsf s}$ is said to be \emph{accepting} if either
one of the following conditions is verified:
\begin{inparaenum}[$(i)$]
 \item $q \in \atmforallstateset$ and 
$c'$ is accepting for all $c' \in \atmnextconf{c}$ or
 \item $q \in \atmexiststateset$ and there exists $c' \in \atmnextconf{c}$ such that $c'$ is accepting.
\end{inparaenum}
Notice that an universal (existential) state always accepts (rejects) if $\atmnextconf{c} = \emptyset$.
A \lbatm \emph{accepts on an initial  input tape }  $\atmtapeconf{s_0}$,  
if the initial configuration 
$\atmconf{\atminitstate}{\atmtapeconf{s_0}}$ is accepting.
\\
\noindent {\bf Hierarchical representation.}
In order to exhibit our encoding proposal, we make use of a hierarchical
representation analogous to the one described in \cite{AY01,LNPP03,LNPP08} for
model checking, and in \cite{MNP2008} for module checking procedures.
Given a finite state machine, the idea of hierarchical representation is to replace two or more substructures
of the machine that are structurally equivalent, by another (structurally equivalent) module, that is a finite state machine itself.
The use of hierarchical representation results in an exponentially
more succinct representation of the system, that amounts (in most cases) to
more efficient model checking procedures (in the other cases, this
does not yield a more efficient behavior, as the analysis requires
a flattening of the machine itself, thus incurring in an exponential blow up in its size).

In our context, this idea can be suitably adapted to deal with the presence
of resources, as follows.
Modules do not represent structurally equivalent substructures,
but substructures that have the same impact on the values of resource variables.
In principle, whenever the analysis is focused on the evolution of
resource variables, it makes sense to consider as equivalent two substructures
that can possibly differ in their structure but whose effect on the set of
resource variables is exactly the same. 
This approach could be thought of as a hierarchical representation based on
\emph{functional} equivalence between substructures, as opposed to the
classical notion of hierarchical representation based on
\emph{structural} equivalence.
\subsection{A reduction from the acceptance problem for \texorpdfstring{\lbatm}{LBATM}}

Given an \lbatm $\mathcal A$ and an input tape
configuration $\atmtapeconf{s_0}$,
we provide a priced game structure $G_{\mathcal A, \atmtapeconf{s_0}}$, with two agents $ag_1$ and $ag_2$,
and a formula $\phi_{\mathcal A, \mathsf{s_0}}$ such that $G_{\mathcal A, \mathsf{s_0}} \models \phi_{\mathcal A, \mathsf{s_0}}$
if and only if $\mathcal A$ accepts on $\atmtapeconf{s_0}$.

In the following, we exhibit the game structure by using a graphical (hierarchical) representation
(Figures \ref{fig:encoding_instr}-\ref{fig:module_shift_right3} in Appendix).
Notice that only significant information is explicitly shown in the pictures.
In particular, labels on transitions (arcs) represent consumptions/productions of resources due to
the execution of the joint move (proponent and opponent moves) associated to that transition.
For example, the label ``$-1i, +1\overline i, +10\mu_L, -10\overline{\mu_L}$'' on the loop transition
of Figure \ref{fig:times10} means that the actions associated to the transition will consume 1 unit of
the (type) resource $i$ and 10 unit of $\overline{\mu_L}$, and will produce 1 unit of
the resource $\overline i$ and 10 unit of $\mu_L$.
Availability of other resources is unchanged, then the relative information is omitted.

The reduction uses the three resource variables $\mu_L$, $\mu$, and $\mu_R$ to encode
the tape configuration, plus  three auxiliary resource variables $i$, $r$, and $t$,
that will be useful during the construction. Moreover, we associate to
the above set of variables the set of \emph{counterbalanced variables}
$\{ \overline{\mu_L}, \overline{\mu}, \overline{\mu_R},
\overline{i}, \overline{r}, \overline{t} \}$.
The idea behind the use of counterbalanced variables, that is also the key idea of
the reduction, consists of designing the game structure in a way that
to every consumption (resp., production) of a resource, say for instance $\mu$,
a corresponding production (resp., consumption) of its
counterbalanced $\overline{\mu}$ exists.
In particular, this is true inside each module of the hierarchical structure, thus the sum of the availability of
a resource variable and its counterbalanced variable is kept constant along all the computation
at every module's entry and exit points, equal to a value $Max$, which depends on
the input of the \lbatm.
This will allow us to force the execution of specific transitions at specific availabilities
of resource variables.
Consider, for example, the node of Figure \ref{fig:times10} with 2 outgoing transitions, one of which is
a loop transition.
The presence of 2 outgoing transitions means that either the proponent or the opponent can choose
between 2 moves. But such a freedom is only potential, as in any moment of the computation the
choice of the next move by the proponent/opponent is constrained by the resource availability:
if the loop transition is enabled, then the availability of the resource $i$ is greater than 0,
and thus the availability of its counterbalanced variable $\overline i$ is less than $Max$, that
means that the other transition, which consumes $Max$ units of the resource $\overline i$, is disabled.
On the contrary, if the non-loop transition is enabled, there are $Max$ units of the resource
$\overline i$ available, and thus the availability of the resource $i$ is 0, that means that the loop
transition is disabled.
Thus, by taking advantage of the features of counterbalanced variables,
we are able to force the executions to have a somehow deterministic behavior.

\medskip

\noindent{\bf Encoding of the tape.}
Without loss of generality, we consider \lbatm on input alphabet
$\atminputsym = \{ \mathsf 1, \mathsf 2, \mathsf B \}$, thus \atmtapesym is the
set $\{ \mathsf 1, \mathsf 2, \mathsf B, \llcorner, \lrcorner \}$.
Recall that the symbols $\mathsf B$, $\llcorner$, and $\lrcorner$
denote the `blank' symbol, the left delimiter, and the right delimiter, respectively.
Tape symbols are encoded
by the digits $0, 1, 2, 3$ and $4$, in a pretty natural way:
$0$ encodes the `blank' symbol,
$1$ and $2$ encode the input symbols $\mathsf 1$ and $\mathsf 2$,
and $3$ and $4$ encode the left and right delimiters.
The tape configuration is encoded by means of the three resource variables $\mu_L$,
$\mu$, and $\mu_R$.
The value of $\mu$ ranges over the set $\{ 0,1,2,3,4 \}$ and encodes the
value stored in the cell currently read by the head (according to the above
encoding of tape symbols into digits).
The value of $\mu_L$ encodes the tape configuration at the left of the
current head position in a forward fashion.
The value of $\mu_R$ encodes the tape configuration at the right of the
current head position in a reverse fashion, that is,
$\mu_R$ encodes the reverse of the string corresponding to the tape configuration at
the right of current head position.
As an example, consider the tape configuration
$\mathsf s = \llcorner \mathsf{B112\underline{1}1B2BB} \lrcorner$,
the symbol read by the head is the underlined one.
Such a configuration is encoded by means of the three resource variables
as follows: $\mu_L = 30112$,
$\mu_L = 1$, and $\mu_R = 400201$.
It can be noticed that the length of the representation of the three
variables $\mu_L$, $\mu$, and $\mu_R$ is proportional to the length
of the tape configuration which is at most linear in the size of the input, 
namely $O(| \atmtapeconf{s_0} |)$.
Using such an encoding, the machine operation ``shift the head to the left''
can be represented by means of the following operations
on resource variables:

\begin{compactitem}
 \item the new value of $\mu_R$ is $\mu_R*10 + \mu$
 \item the new value of $\mu$ is $\mu_L \mod 10$,
 \item the new value of $\mu_L$ is $\mu_L/10$ ( $/$ is the integer division),
\end{compactitem}
The operation ``shift the head to the right'' can be encoded analogously.

Notice that in order to encode in polynomial time
the operations of shifting the head to left and right, we encode the string to the
right of the current head position in a reverse order.
Indeed, in this way the symbol stored on the cell immediately to the right
of the head corresponds to the least significant digit of $\mu_R$,
and thus can be accessed by using the module operation ($\mu_R \mod 10$).

\medskip

\noindent{\bf Encoding of the instructions.}
The encoding of the instructions is depicted in \figurename~\ref{fig:encoding_instr}.
Transitions starting from a node labeled
$\fullstate{\atmstate q}{\lambda}$ represent all the possible instructions
matching the full state $\fullstate{\atmstate q}{\lambda}$ of the \lbatm,
that is, all the instructions that can be possibly performed
at the full state $\fullstate{\atmstate q}{\lambda}$.

More in detail, given a full state $\fullstate{\atmstate q}{\lambda}$
of the machine, with $\atmstate q \in \atmexiststateset$, the encoding of
the set
$ \{ \atminstr{\fullstate{\atmstate q}{\lambda}}{\atmaction{\atmstate{r_1}}{\nu_1}{\sim_1}},
\atminstr{\fullstate{\atmstate q}{\lambda}}{\atmaction{\atmstate{r_2}}{\nu_2}{\sim_2}}, \ldots,
\atminstr{\fullstate{\atmstate q}{\lambda}}{\atmaction{\atmstate{r_m}}{\nu_m}{\sim_m}} \} $
of matching instructions
is shown in \figurename~\ref{fig:encoding_ATM_exists_instr}, (recall that $\sim_i \in \{\leftarrow, \rightarrow\nobreak \})$.
Analogously, the encoding of the set of instructions matching the
full state $\fullstate{\atmstate q}{\lambda}$, with $\atmstate q \in \atmforallstateset$,
is shown in \figurename~\ref{fig:encoding_ATM_forall_instr}.
Let us underline that the 
action profiles $\langle \alpha_1, \beta \rangle, \ldots, \langle \alpha_m, \beta \rangle$ labeling 
transitions corresponding to an existential state are such that the first agent $ag_1$
has the capability to force a specific transition (instruction) to be executed, depending on the
choice of the $\alpha_i$ for the next action, independently from the choice $\beta$ of the other agent $ag_2$.
On the other hand, the action profiles $\langle \alpha, \beta_1 \rangle, \ldots, \langle \alpha, \beta_m \rangle$
labeling transitions corresponding to an universal state are such that the roles of the agents are exchanged.


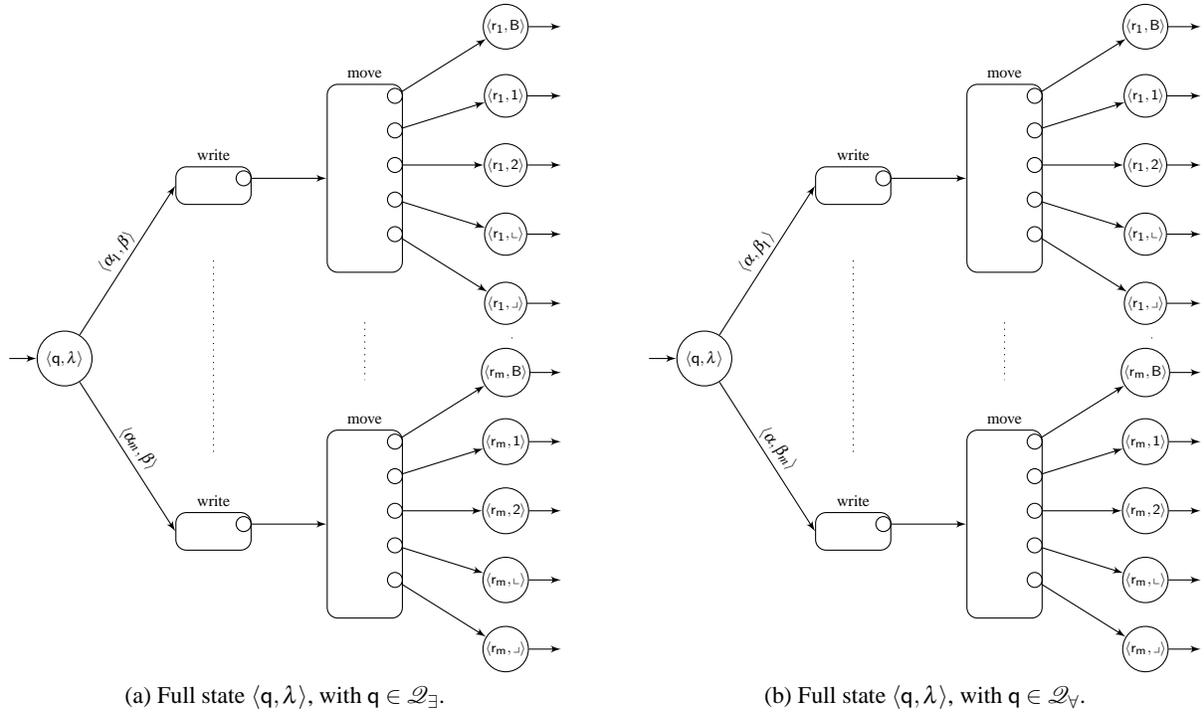
\begin{figure}[t!]
\tiny
\centering

\subfloat[Full state \fullstate{\atmstate q}{\lambda}, with $\atmstate q \in \atmexiststateset$.]{
\label{fig:encoding_ATM_exists_instr}
\begin{tikzpicture}[scale=.46,xscale=.8]

 \draw (0,0) node(q_lambda)[circle,draw]{\fullstate{\atmstate q}{\lambda}};
 \draw[-latex'] (q_lambda) ++(-2,0) -- (q_lambda);

\mybox{(q_lambda) ++(4,5)}{write1}{
\begin{minipage}{5cm}
\end{minipage}
}{write}{1}

\mybox{(q_lambda) ++(4,-5)}{writem}{
\begin{minipage}{5cm}
\end{minipage}
}{write}{1}

\draw (write1) ++(0,-2)node(write1_south){};
\draw (writem) ++(0,2)node(writem_north){};
\draw[dotted] (write1_south) -- (writem_north);

\mybox{(write1_out1) ++(3,0)}{move1}{
\begin{minipage}{5cm}
\end{minipage}
}{move}{5}

\mybox{(writem_out1) ++(3,0)}{movem}{
\begin{minipage}{5cm}
\end{minipage}
}{move}{5}

\draw (move1) ++(0,-4)node(move1_south){};
\draw (movem) ++(0,4)node(movem_north){};
\draw[dotted] (move1_south) -- (movem_north);

 \draw (move1_out1) ++(4,2) node[inner sep=.5pt,scale=.9](r1_blank)[circle,draw]{\fullstate{\atmstate{r_1}}{\mathsf{B}}};
 \draw (move1_out2) ++(4,1) node[inner sep=.5pt,scale=.9](r1_1)[circle,draw]{\fullstate{\atmstate{r_1}}{\mathsf{1}}};
 \draw (move1_out3) ++(4,0) node[inner sep=.5pt,scale=.9](r1_2)[circle,draw]{\fullstate{\atmstate{r_1}}{\mathsf{2}}};
 \draw (move1_out4) ++(4,-1) node[inner sep=.5pt,scale=.9](r1_left)[circle,draw]{\fullstate{\atmstate{r_1}}{\llcorner}};
 \draw (move1_out5) ++(4,-2) node[inner sep=.5pt,scale=.9](r1_right)[circle,draw]{\fullstate{\atmstate{r_1}}{\lrcorner}};

\draw[-latex'] (q_lambda) --node[sloped,above]{$\langle \alpha_1, \beta \rangle$} (write1.west);
\draw[-latex'] (write1_out1) -- (move1);

\draw[-latex'] (move1_out1) -- (r1_blank);
\draw[-latex'] (move1_out2) -- (r1_1);
\draw[-latex'] (move1_out3) -- (r1_2);
\draw[-latex'] (move1_out4) -- (r1_left);
\draw[-latex'] (move1_out5) -- (r1_right);

\draw[-latex'] (r1_blank) -- ++(2,0);
\draw[-latex'] (r1_1) -- ++(2,0);
\draw[-latex'] (r1_2) -- ++(2,0);
\draw[-latex'] (r1_left) -- ++(2,0);
\draw[-latex'] (r1_right) -- ++(2,0);

 \draw (movem_out1) ++(4,2) node[inner sep=.5pt,scale=.9](rm_blank)[circle,draw]{\fullstate{\atmstate{r_m}}{\mathsf{B}}};
 \draw (movem_out2) ++(4,1) node[inner sep=.5pt,scale=.9](rm_1)[circle,draw]{\fullstate{\atmstate{r_m}}{\mathsf{1}}};
 \draw (movem_out3) ++(4,0) node[inner sep=.5pt,scale=.9](rm_2)[circle,draw]{\fullstate{\atmstate{r_m}}{\mathsf{2}}};
 \draw (movem_out4) ++(4,-1) node[inner sep=.5pt,scale=.9](rm_left)[circle,draw]{\fullstate{\atmstate{r_m}}{\llcorner}};
 \draw (movem_out5) ++(4,-2) node[inner sep=.5pt,scale=.9](rm_right)[circle,draw]{\fullstate{\atmstate{r_m}}{\lrcorner}};

\draw[-latex'] (q_lambda) --node[sloped,above]{$\langle \alpha_m, \beta \rangle$} (writem.west);
\draw[-latex'] (writem_out1) -- (movem);

\draw[-latex'] (movem_out1) -- (rm_blank);
\draw[-latex'] (movem_out2) -- (rm_1);
\draw[-latex'] (movem_out3) -- (rm_2);
\draw[-latex'] (movem_out4) -- (rm_left);
\draw[-latex'] (movem_out5) -- (rm_right);

\draw[-latex'] (rm_blank) -- ++(2,0);
\draw[-latex'] (rm_1) -- ++(2,0);
\draw[-latex'] (rm_2) -- ++(2,0);
\draw[-latex'] (rm_left) -- ++(2,0);
\draw[-latex'] (rm_right) -- ++(2,0);

\draw (r1_right) ++(0,-1)node(r1_right_north){};
\draw (rm_blank) ++(0,1)node(rm_blank_south){};
\draw[dotted] (r1_right_north) -- (rm_blank_south);

\end{tikzpicture}
}
\hspace{10mm}
\subfloat[Full state \fullstate{\atmstate q}{\lambda}, with $\atmstate q \in \atmforallstateset$.]{
\label{fig:encoding_ATM_forall_instr}
\begin{tikzpicture}[scale=.46,xscale=.8]

 \draw (0,0) node(q_lambda)[circle,draw]{\fullstate{\atmstate q}{\lambda}};
 \draw[-latex'] (q_lambda) ++(-2,0) -- (q_lambda);

\mybox{(q_lambda) ++(4,5)}{write1}{
\begin{minipage}{5cm}
\end{minipage}
}{write}{1}

\mybox{(q_lambda) ++(4,-5)}{writem}{
\begin{minipage}{5cm}
\end{minipage}
}{write}{1}

\draw (write1) ++(0,-2)node(write1_south){};
\draw (writem) ++(0,2)node(writem_north){};
\draw[dotted] (write1_south) -- (writem_north);

\mybox{(write1_out1) ++(3,0)}{move1}{
\begin{minipage}{5cm}
\end{minipage}
}{move}{5}

\mybox{(writem_out1) ++(3,0)}{movem}{
\begin{minipage}{5cm}
\end{minipage}
}{move}{5}

\draw (move1) ++(0,-4)node(move1_south){};
\draw (movem) ++(0,4)node(movem_north){};
\draw[dotted] (move1_south) -- (movem_north);

 \draw (move1_out1) ++(4,2) node[inner sep=.5pt,scale=.9](r1_blank)[circle,draw]{\fullstate{\atmstate{r_1}}{\mathsf{B}}};
 \draw (move1_out2) ++(4,1) node[inner sep=.5pt,scale=.9](r1_1)[circle,draw]{\fullstate{\atmstate{r_1}}{\mathsf{1}}};
 \draw (move1_out3) ++(4,0) node[inner sep=.5pt,scale=.9](r1_2)[circle,draw]{\fullstate{\atmstate{r_1}}{\mathsf{2}}};
 \draw (move1_out4) ++(4,-1) node[inner sep=.5pt,scale=.9](r1_left)[circle,draw]{\fullstate{\atmstate{r_1}}{\llcorner}};
 \draw (move1_out5) ++(4,-2) node[inner sep=.5pt,scale=.9](r1_right)[circle,draw]{\fullstate{\atmstate{r_1}}{\lrcorner}};

\draw[-latex'] (q_lambda) --node[sloped,above]{$\langle \alpha, \beta_1 \rangle$} (write1.west);
\draw[-latex'] (write1_out1) -- (move1);

\draw[-latex'] (move1_out1) -- (r1_blank);
\draw[-latex'] (move1_out2) -- (r1_1);
\draw[-latex'] (move1_out3) -- (r1_2);
\draw[-latex'] (move1_out4) -- (r1_left);
\draw[-latex'] (move1_out5) -- (r1_right);

\draw[-latex'] (r1_blank) -- ++(2,0);
\draw[-latex'] (r1_1) -- ++(2,0);
\draw[-latex'] (r1_2) -- ++(2,0);
\draw[-latex'] (r1_left) -- ++(2,0);
\draw[-latex'] (r1_right) -- ++(2,0);

 \draw (movem_out1) ++(4,2) node[inner sep=.5pt,scale=.9](rm_blank)[circle,draw]{\fullstate{\atmstate{r_m}}{\mathsf{B}}};
 \draw (movem_out2) ++(4,1) node[inner sep=.5pt,scale=.9](rm_1)[circle,draw]{\fullstate{\atmstate{r_m}}{\mathsf{1}}};
 \draw (movem_out3) ++(4,0) node[inner sep=.5pt,scale=.9](rm_2)[circle,draw]{\fullstate{\atmstate{r_m}}{\mathsf{2}}};
 \draw (movem_out4) ++(4,-1) node[inner sep=.5pt,scale=.9](rm_left)[circle,draw]{\fullstate{\atmstate{r_m}}{\llcorner}};
 \draw (movem_out5) ++(4,-2) node[inner sep=.5pt,scale=.9](rm_right)[circle,draw]{\fullstate{\atmstate{r_m}}{\lrcorner}};

\draw[-latex'] (q_lambda) --node[sloped,above]{$\langle \alpha, \beta_m \rangle$} (writem.west);
\draw[-latex'] (writem_out1) -- (movem);

\draw[-latex'] (movem_out1) -- (rm_blank);
\draw[-latex'] (movem_out2) -- (rm_1);
\draw[-latex'] (movem_out3) -- (rm_2);
\draw[-latex'] (movem_out4) -- (rm_left);
\draw[-latex'] (movem_out5) -- (rm_right);

\draw[-latex'] (rm_blank) -- ++(2,0);
\draw[-latex'] (rm_1) -- ++(2,0);
\draw[-latex'] (rm_2) -- ++(2,0);
\draw[-latex'] (rm_left) -- ++(2,0);
\draw[-latex'] (rm_right) -- ++(2,0);

\draw (r1_right) ++(0,-1)node(r1_right_north){};
\draw (rm_blank) ++(0,1)node(rm_blank_south){};
\draw[dotted] (r1_right_north) -- (rm_blank_south);

\end{tikzpicture}
}

\caption{Encoding of the set of instructions matching a full state \fullstate{\atmstate q}{\lambda} of a \lbatm.}
\label{fig:encoding_instr}
\end{figure}

The \lbatm representation of \figurename~\ref{fig:encoding_instr}
is hierarchical and involves the modules \emph{write} and \emph{move}.
The former encodes the rewriting of the head cell performed by $\mathcal A$ and, to this aim,
makes use of one of the following modules (\figurename~\ref{fig:inc_dec}),
depending on the symbol $\lambda$ read by the head,
and on the symbol $\nu$ to be written:
\begin{compactitem}
 \item \emph{inc}, depicted in \figurename~\ref{fig:inc},
is used when the rewriting corresponds to an increment, for example, when the symbol $\mathsf 2$ has to
be written in place of the symbol $\mathsf 1$;
 \item \emph{double\_inc}, depicted in \figurename~\ref{fig:double_inc},
is used when the rewriting corresponds to a double increment,
for example, when the symbol $\mathsf 2$ (encoded as $2$) has to
be written in place of the symbol $\mathsf B$ (encoded as $0$);
 \item \emph{dec}, depicted in \figurename~\ref{fig:dec},
is used when the rewriting corresponds to a decrement, for example, when the symbol $\mathsf 1$ has to
be written in place of the symbol $\mathsf 2$;
 \item \emph{double\_dec}, depicted in \figurename~\ref{fig:double_dec},
is used when the rewriting corresponds to a double decrement, for example, when the symbol $\mathsf B$ has to
be written in place of the symbol $\mathsf 2$.
\end{compactitem}
Obviously, the module does nothing when the symbol to be written corresponds
to the symbol currently stored in the head cell.


\begin{figure}[b!]
\tiny
\centering
\subfloat[Module \emph{inc}.]{
\label{fig:inc}
\begin{tikzpicture}[scale=.5]
\draw(-1,.5) (3,-.5);
 \draw (0,0) node(n1)[circle,draw]{} ++ (2,0) node(n2)[circle,draw]{};
 \draw[-latex'] (n1) ++(-1,0) -- (n1);
 \draw[-latex'] (n1) -- node[above]{$+1 \mu,-1\overline{\mu}$} (n2);
 \draw[-latex'] (n2) -- ++(1,0);
\end{tikzpicture}
}
\hspace{1cm}
\subfloat[Module \emph{double\_inc}.]{
\label{fig:double_inc}
\begin{tikzpicture}[scale=.5]
\draw(-.75,.5) (5.75,-.5);

  \mybox{(0,0)}{inc1}{
    \begin{minipage}{5cm}
    \end{minipage}
  }{\emph{inc}}{1}

 \draw[-latex'] (inc1) ++(-1.75,0) -- (inc1);

  \mybox{(inc1) ++ (2,0)}{inc2}{
    \begin{minipage}{5cm}
    \end{minipage}
  }{\emph{inc}}{1}

  \draw[-latex'] (inc1) -- (inc2);

 \draw[-latex'] (inc2) -- ++(1.75,0);
\end{tikzpicture}
}
\hspace{1cm}
\subfloat[Module \emph{dec}.]{
\label{fig:dec}
\begin{tikzpicture}[scale=.5]
\draw(-1,.5) (3,-.5);
 \draw (0,0) node(n1)[circle,draw]{} ++ (2,0) node(n2)[circle,draw]{};
 \draw[-latex'] (n1) ++(-1,0) -- (n1);
 \draw[-latex'] (n1) -- node[above]{$-1 \mu,+1\overline{\mu}$} (n2);
 \draw[-latex'] (n2) -- ++(1,0);
\end{tikzpicture}
}
\hspace{1cm}
\subfloat[Module \emph{double\_dec}.]{
\label{fig:double_dec}
\begin{tikzpicture}[scale=.5]
\draw(-.75,.5) (5.75,-.5);

  \mybox{(0,0)}{dec1}{
    \begin{minipage}{5cm}
    \end{minipage}
  }{\emph{dec}}{1}

 \draw[-latex'] (dec1) ++(-1.75,0) -- (dec1);

  \mybox{(dec1) ++ (2,0)}{dec2}{
    \begin{minipage}{5cm}
    \end{minipage}
  }{\emph{dec}}{1}

  \draw[-latex'] (dec1) -- (dec2);
 \draw[-latex'] (dec2) -- ++(1.75,0);

\end{tikzpicture}
}

\caption{Encoding of the module \emph{write}.}
\label{fig:inc_dec}
\end{figure}
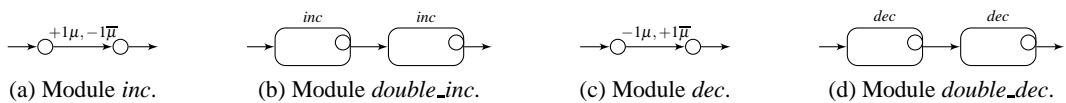

The module \emph{move} encodes the shift (to right or to left) of the head.
It is designed in a way that the only next location that can be reached by the
game is consistent with the value stored on the new head cell (after the shift operation).
In \figurename~\ref{fig:module_shift_right1} and \ref{fig:module_shift_right2} the sub-modules encoding the operation
``shift to right'' are depicted.
The encoding of the operation ``shift to left'' can be realized analogously.


\begin{figure}[t!]
\tiny
\centering
\subfloat[Module \emph{shift\_right}.]{
\label{fig:submodule_shift_right}
\begin{tikzpicture}[scale=.5,xscale=.8]
  \draw (0,0)node(times10temp){};
  \mybox{(0,0)}{times10}{
    \begin{minipage}{10mm}
      \hspace{0cm}
    \end{minipage}
  }{\emph{times\_10($\mu_L$)}}{1}
  \draw[-latex'] (times10) ++(-2.5,0) -- (times10);

  \mybox{(times10) ++(3.8,0)}{add}{
    \begin{minipage}{10mm}
      \hspace{0cm}
    \end{minipage}
  }{\emph{add($\mu_L$,$\mu$)}}{1}

  \draw[-latex'] (times10) -- (add);

  \mybox{(times10temp) ++(.5,-3)}{div10}{
    \begin{minipage}{10mm}
      \hspace{0cm}
    \end{minipage}
  }{\emph{div\_10($\mu_R$)}}{1}

  \draw (add.south) ++(-1,-.5)node(c1){} (div10.west) ++(-1,1)node(c2){};
  \draw plot [smooth, tension=.5] coordinates {
  (add.south) (c1) (c2) (div10.west) };
  \draw[-latex'] (div10.west) ++(-.1,.05) -- (div10.west);

  \mybox{(div10) ++(3.8,0)}{assign}{
    \begin{minipage}{10mm}
      \hspace{0cm}
    \end{minipage}
  }{\emph{assign($\mu$,$r$)}}{1}

  \draw[-latex'] (div10) -- (assign);

  \draw (assign) ++(3.8,0)node(chooseTemp){};
  \mybox{(assign) ++(3.8,1)}{choose}{
    \begin{minipage}{10mm}
      \hspace{0cm}
    \end{minipage}
  }{\emph{choose\_next\_state($\mu$)}}{5}

  \draw[-latex'] (assign) -- (chooseTemp);

  \draw[-latex'] (choose_out1) -- ++(1,0);
  \draw[-latex'] (choose_out2) -- ++(1,0);
  \draw[-latex'] (choose_out3) -- ++(1,0);
  \draw[-latex'] (choose_out4) -- ++(1,0);
  \draw[-latex'] (choose_out5) -- ++(1,0);

\end{tikzpicture}
}
\hspace{1cm}
\subfloat[Module \emph{times\_10($\mu_L$)}.]{
\label{fig:times10}
\begin{tikzpicture}[scale=.5,xscale=.8]
\draw(0,-1) ++(0,-1.5);
  \mybox{(0,0)}{assign}{
    \begin{minipage}{15mm}
      \hspace{0cm}
    \end{minipage}
  }{\emph{assign($i$,$\mu_L$)}}{1}

  \draw[-latex'] (assign) ++(-3,0) -- (assign);

  \mybox{(assign) ++ (4,0)}{tozero}{
    \begin{minipage}{15mm}
      \hspace{0cm}
    \end{minipage}
  }{\emph{to\_zero($\mu_L$)}}{1}

  \draw[-latex'] (assign) -- (tozero);

  \draw (tozero) ++(4,0)node(loop)[draw,circle]{};
  \draw[-latex'] (tozero) -- (loop);

  \draw[-latex'] (loop.north) .. controls ++(0,1.5) and ++(2.5,1.5) ..
    node(a)[right]{
      \begin{tabular}{l}
        $-1i, + 1 \overline i$ \\
        $+10 \mu_L, -10 \overline{\mu_L}$
      \end{tabular}} (loop.east);

  \draw (loop) ++(2,-2)node(node1)[draw,circle]{};
  \draw[-latex'] (loop.south) .. controls ++(.5,-1.5) .. node[pos=.75,above=3mm]{$-Max \ \overline{i}$} (node1);

  \draw (node1) ++(2,0)node(node2)[draw,circle]{};
  \draw[-latex'] (node1) --node[above]{$+Max \ \overline{i}$} (node2);

  \draw[-latex'] (node2) -- ++(1,0);

\end{tikzpicture}
}

\subfloat[Module \emph{assign($x_1$, $x_2$)}.]{
\label{fig:assign}
\begin{tikzpicture}[scale=.5,xscale=.8]
\draw(-1,2) ++(18,-6.5);

  \mybox{(0,0)}{tozero}{
    \begin{minipage}{15mm}
      \hspace{0cm}
    \end{minipage}
  }{\emph{to\_zero($x_1$)}}{1}

  \draw[-latex'] (tozero) ++(-3,0) -- (tozero);

  \mybox{(tozero) ++(4,0)}{tozero2}{
    \begin{minipage}{15mm}
      \hspace{0cm}
    \end{minipage}
  }{\emph{to\_zero($t$)}}{1}

  \draw[-latex'] (tozero) -- (tozero2);

  \draw (tozero2) ++(4,0)node(loop)[draw,circle]{};
  \draw[-latex'] (tozero2) -- (loop);

  \draw[-latex'] (loop.north) .. controls ++(0,1.5) and ++(2.5,1.5) ..
    node(a)[right]{
      \begin{tabular}{l}
        $-1 x_2, + 1 \overline{x_2}$ \\
        $+1 x_1, - 1 \overline{x_1}$ \\
        $+1 t, -1 \overline{t}$
      \end{tabular}} (loop.east);

  \draw (loop) ++(-10,-3)node(node1)[draw,circle]{};

  \draw (loop.south)++(0,-1)node(c1){} (node1.west)++(0,1)node(c2){};
  \draw (c2)node[above]{$-Max \ \overline{x_2}$};
  \draw plot [smooth, tension=.5] coordinates {
  (loop.south) (c1) (c2) (node1.west) };
  \draw[-latex'] (node1.west) ++(-.001,.0007) -- (node1.west);

  \draw (node1) ++(4,0)node(node2)[draw,circle]{};
  \draw[-latex'] (node1) --node[above]{$+Max \ \overline{x_2}$} (node2);

  \draw[-latex'] (node2.south west) .. controls ++(0,-1.5) and ++(2.5,-1.5) ..
    node[pos=.4,left]{
      \begin{tabular}{r}
        $+1 x_2, -1 \overline{x_2}$ \\
        $-1 t, +1 \overline{t}$
      \end{tabular}} (node2.south east);

  \draw (node2) ++(4,0)node(node3)[draw,circle]{};
  \draw[-latex'] (node2) --node[above]{$-Max \ \overline{t}$} (node3);

  \draw (node3) ++(4,0)node(node4)[draw,circle]{};
  \draw[-latex'] (node3) --node[above]{$+Max \ \overline{t}$} (node4);

  \draw[-latex'] (node4) -- ++(1,0);
\end{tikzpicture}
}
%
%
\subfloat[Module \emph{to\_zero($x$)}.]{
\label{fig:tozero}
\begin{tikzpicture}[scale=.5,xscale=.8]
\draw(-1.5,1.5) (9,-3);

  \draw (0,0)node(loop)[draw,circle]{};
  \draw[-latex'] (loop) ++(-1,0) -- (loop);

  \draw[-latex'] (loop.north west) .. controls ++(0,1.5) and ++(2.5,1.5) ..
    node[left]{
      \begin{tabular}{l}
        $-1 x, + 1 \overline{x}$
      \end{tabular}} (loop.north east);

  \draw (loop) ++(4,0)node(node)[draw,circle]{};
  \draw[-latex'] (loop) --node[above]{$-Max \ \overline{x}$} (node);

  \draw (node) ++(4,0)node(node1)[draw,circle]{};
  \draw[-latex'] (node) --node[above]{$+Max \ \overline{x}$} (node1);

  \draw[-latex'] (node1) -- ++(1,0);
\end{tikzpicture}
}

\caption{Encoding of the module \emph{shift\_right} - part I.}
\label{fig:module_shift_right1}
\end{figure}
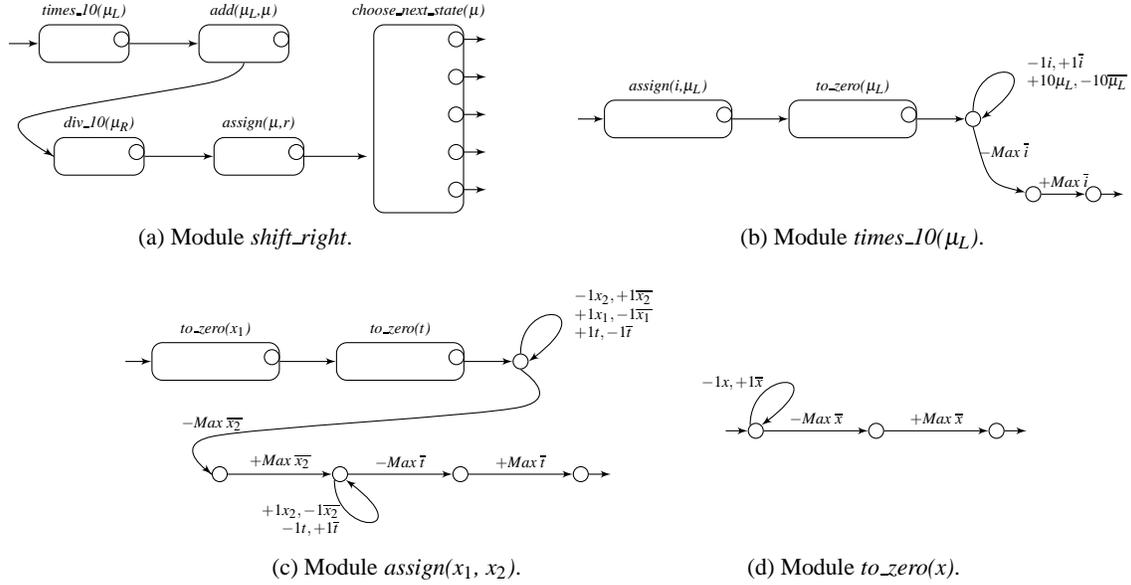

As an example, we describe the first two modules of Figure \ref{fig:module_shift_right1}.
The module $shift\_right$, depicted in Figure \ref{fig:submodule_shift_right},
is performed through the following steps:
\begin{compactitem}
\item multiply by 10 the value of $\mu_L$ (module $times\_10(\mu_L)$),
\item increment the value of $\mu_L$ by the value of $\mu$ (module $add(\mu_L,\mu)$),
\item divide by 10 the value of $\mu_R$ (module $div\_10(\mu_R)$ --- the remainder of
the division is stored in the resource variable $r$),
\item assign to the resource variable $\mu$ the value of $r$ (module $assign(\mu,r)$),
\item suitably lead the computation to the location corresponding to the next state of the
\lbatm, depending on the value read by the head, that is, the value stored on the resource variable $\mu$
(module $choose\_next\_state(\mu)$).
\end{compactitem}
The module $times\_10(\mu_L)$, that multiplies by 10 the value of $\mu_L$ (Figure \ref{fig:times10}), is performed
by storing the value of $\mu_L$ in the resource variable $i$, by setting the value of $\mu_L$ to 0,
and then by executing a transition (the loop transition), which consumes 1 unit of $i$ and produces
10 units of $\mu_L$
(the suitable quantity of the counterbalanced variables is produced or consumed as well, to keep the sum
constant) as long as items of the resource $i$ are available.
When the availability of $i$ goes down to 0, the other transition is executed (the last transition
is needed to keep constant the sum between $i$ and its counterbalanced variable $\overline i$).
It is easy to convince oneself that the value of $\mu_L$ in the exit node is equal to its value
in the entry node times 10, and that the sum of each variable and its counterbalanced one is constant.
As a last remark, we point out that the names of some of the modules are parametric, in the sense that
the arguments between parenthesis are not actual resource variables, but parameters (e.g., $x$, $x_1$, $x_2$)
to be instantiated. We adopted this notation for modules that are used more than once,
and that are instantiated with actual resource variables when they are used (e.g., the module
$assign$ depicted in Figure \ref{fig:assign} is called $assign(x_1,x_2)$ and it is used, for instance,
inside the module $times\_10(\mu_L)$ (Figure \ref{fig:times10}), where $x_1$ (resp., $x_2$) is instantiated with
$i$ (resp., $\mu_L$), and inside the module $add(\mu_L,\mu)$ (Figure \ref{fig:add}), where $x_1$ (resp., $x_2$)
is instantiated with $t$ (resp., $\mu$).


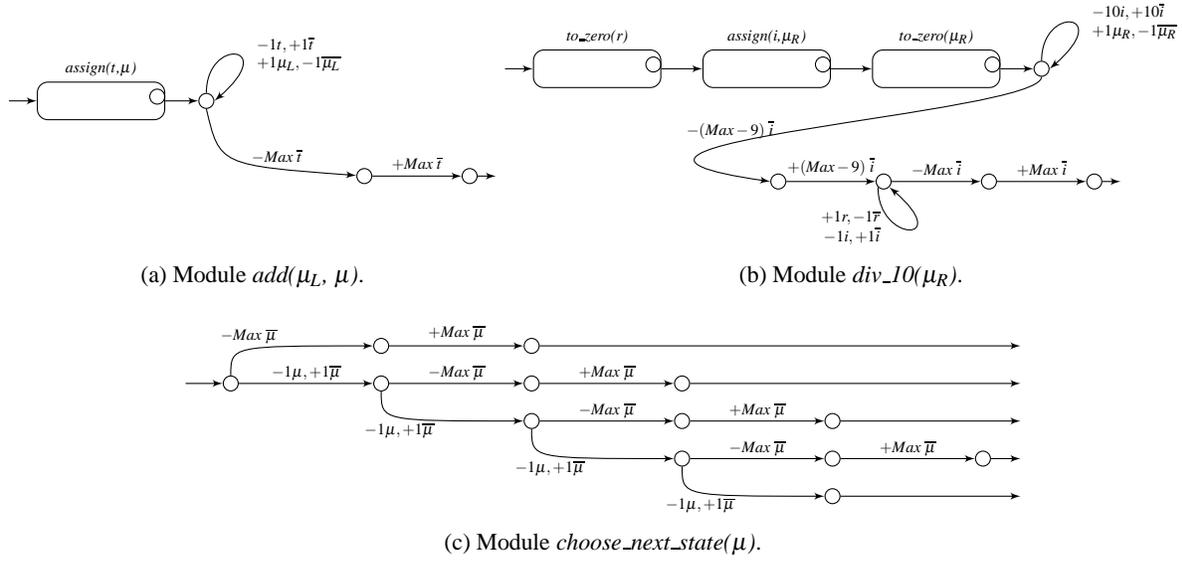
\begin{figure}[t]
\tiny
\centering

\subfloat[Module \emph{add($\mu_L$, $\mu$)}.]{
\label{fig:add}
\begin{tikzpicture}[scale=.5,xscale=.7]
\draw(0,-1) ++(0,-3);
  \mybox{(0,0)}{assign}{
    \begin{minipage}{15mm}
      \hspace{0cm}
    \end{minipage}
  }{\emph{assign($t$,$\mu$)}}{1}

  \draw[-latex'] (assign) ++(-3.5,0) -- (assign);

  \draw (assign) ++(4,0)node(loop)[draw,circle]{};
  \draw[-latex'] (assign) -- (loop);

  \draw[-latex'] (loop.north) .. controls ++(0,1.5) and ++(2.5,1.5) ..
    node(a)[right]{
      \begin{tabular}{l}
        $-1 t, + 1 \overline {t}$ \\
        $+1 \mu_L, -1 \overline{\mu_L}$
      \end{tabular}} (loop.east);

  \draw (loop) ++(6,-2)node(node1)[draw,circle]{};
  \draw[-latex'] (loop.south) .. controls ++(.5,-1.5) .. node[pos=.75,above]{$-Max \ \overline{t}$} (node1);

  \draw (node1) ++(4,0)node(node2)[draw,circle]{};
  \draw[-latex'] (node1) --node[above]{$+Max \ \overline{t}$} (node2);

  \draw[-latex'] (node2) -- ++(1,0);

\end{tikzpicture}
}
\subfloat[Module \emph{div\_10($\mu_R$)}.]{
\label{fig:div10}
\begin{tikzpicture}[scale=.5,xscale=.7]
\draw(0,-1);

  \mybox{(0,0)}{tozero}{
    \begin{minipage}{15mm}
      \hspace{0cm}
    \end{minipage}
  }{\emph{to\_zero($r$)}}{1}

  \draw[-latex'] (tozero) ++(-3.5,0) -- (tozero);

  \mybox{(tozero) ++(4,0)}{assign}{
    \begin{minipage}{15mm}
      \hspace{0cm}
    \end{minipage}
  }{\emph{assign($i$,$\mu_R$)}}{1}

  \draw[-latex'] (tozero) -- (assign);

  \mybox{(assign) ++ (4,0)}{tozero}{
    \begin{minipage}{15mm}
      \hspace{0cm}
    \end{minipage}
  }{\emph{to\_zero($\mu_R$)}}{1}

  \draw[-latex'] (assign) -- (tozero);

  \draw (tozero) ++(4,0)node(loop)[draw,circle]{};
  \draw[-latex'] (tozero) -- (loop);

  \draw[-latex'] (loop.north) .. controls ++(0,1.5) and ++(2.5,1.5) ..
    node(a)[right]{
      \begin{tabular}{l}
        $-10i, + 10 \overline i$ \\
        $+1 \mu_R, -1 \overline{\mu_R}$
      \end{tabular}} (loop.east);

  \draw (loop) ++(-10,-3)node(node1)[draw,circle]{};
  \draw[-latex'] (loop.south) .. controls +(down:1cm) and +(-10,1) ..
    node[pos=.65,above]{$-(Max-9) \ \overline{i}$} (node1.west);

  \draw (node1) ++(4,0)node(node2)[draw,circle]{};
  \draw[-latex'] (node1) --node[above]{$+(Max-9) \ \overline{i}$} (node2);

  \draw[-latex'] (node2.south west) .. controls ++(0,-1.5) and ++(2.5,-1.5) ..
    node[pos=.4,left]{
      \begin{tabular}{r}
        $+1 r, -1 \overline r$ \\
        $-1 i, +1 \overline{i}$
      \end{tabular}} (node2.south east);

  \draw (node2) ++(4,0)node(node3)[draw,circle]{};
  \draw[-latex'] (node2) --node[above]{$-Max \ \overline{i}$} (node3);

  \draw (node3) ++(4,0)node(node4)[draw,circle]{};
  \draw[-latex'] (node3) --node[above]{$+Max \ \overline{i}$} (node4);

  \draw[-latex'] (node4) -- ++(1,0);
\end{tikzpicture}
}

\subfloat[Module \emph{choose\_next\_state($\mu$)}.]{
\label{fig:choosenextstate}
\begin{tikzpicture}[scale=.5]

  \draw (0,0)node(a)[draw,circle]{};
  \draw[-latex'] (a) ++(-1.2,0) -- (a);

  \draw (a) ++(4,1)node(b)[draw,circle]{};
  \draw[-latex'] (a) .. controls ++(0,1) ..node[above]{$-Max \ \overline{\mu}$} (b);

  \draw (a) ++(4,0)node(c)[draw,circle]{};
  \draw[-latex'] (a) --node[above=-1]{$-1 \mu, +1 \overline{\mu}$} (c);

  \draw (c) ++(4,0)node(d)[draw,circle]{};
  \draw[-latex'] (c) --node[above=-1]{$-Max \ \overline{\mu}$} (d);

  \draw (b) ++(4,0)node(e)[draw,circle]{};
  \draw[-latex'] (b) --node[above]{$+Max \ \overline{\mu}$} (e);

  \draw (d) ++(4,0)node(f)[draw,circle]{};
  \draw[-latex'] (d) --node[above=-1]{$+Max \ \overline{\mu}$} (f);

  \draw (c) ++(4,-1)node(g)[draw,circle]{};
  \draw[-latex'] (c) .. controls ++(0,-1) ..node[below]{$-1 \mu, +1 \overline{\mu}$} (g);

  \draw (g) ++(4,-1)node(h)[draw,circle]{};
  \draw[-latex'] (g) .. controls ++(0,-1) ..node[below]{$-1 \mu, +1 \overline{\mu}$} (h);

  \draw (g) ++(4,0)node(g1)[draw,circle]{};
  \draw[-latex'] (g) --node[above=-1]{$-Max \ \overline{\mu}$} (g1);

  \draw (g1) ++(4,0)node(g2)[draw,circle]{};
  \draw[-latex'] (g1) --node[above=-1]{$+Max \ \overline{\mu}$} (g2);

  \draw (h) ++(4,-1)node(i)[draw,circle]{};
  \draw[-latex'] (h) .. controls ++(0,-1) ..node[below]{$-1 \mu, +1 \overline{\mu}$} (i);

  \draw (h) ++(4,0)node(h1)[draw,circle]{};
  \draw[-latex'] (h) --node[above=-1]{$-Max \ \overline{\mu}$} (h1);

  \draw (h1) ++(4,0)node(h2)[draw,circle]{};
  \draw[-latex'] (h1) --node[above=-1]{$+Max \ \overline{\mu}$} (h2);

  \draw[-latex'] (e) -- ++(13,0);
  \draw[-latex'] (f) -- ++(9,0);
  \draw[-latex'] (g2) -- ++(5,0);
  \draw[-latex'] (h2) -- ++(1,0);
  \draw[-latex'] (i) -- ++(5,0);

\end{tikzpicture}
}

\caption{Encoding of the module \emph{shift\_right} - part II.}
\label{fig:module_shift_right2}
\end{figure}

Now, as resource productions are involved in the reduction, we need to guarantee
that the availability of each resource never exceeds the initial one.
To this end
the values of the components of the vector $\vec{m_0}$ of initial resource availability are set to
the value $Max = 322 \ldots 224$, that is the largest number corresponding to an encoding of any
tape configuration (precisely, it encodes the tape configuration
$\mathsf{\llcorner 22 \ldots 22 \lrcorner}$).
\ignore {
Thus, since all variables have the same initial availability, namely $Max$, 
this is true for a variable and its counterbalanced variable, as well, and from this a problem arises: 
no matter whether a resource is consumed or produced by a module,
the value of either the variable or its counterbalanced variable, will exceed the
initial value $Max$. For instance, suppose that the first module has the effect of consuming
one item of the resource $\mu$. As modules keep constant the sum of a variable and its
counterbalanced variable, then we have that the module can only exit by setting the availability of
$\overline{\mu}$ to $Max+1$, that means that some illegal transition would have been executed.

To overcome such an issue, }
Before starting  the simulation of the \lbatm, a preliminary step, depicted in
\figurename~\ref{fig:preliminary_step},  modifies the value of the resource
variables in such a way that  they correctly encode the input tape  \atmtapeconf{s_0}  and
 the sum of the availability of each 
 resource variable and its counterbalanced is equal to  $Max$. Thus, the value  of the resource variables never exceed $Max$.


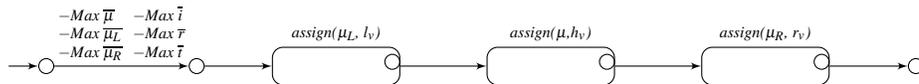
\begin{figure}[b]
\tiny
\begin{tikzpicture}[scale=.5]
\draw(0,-1);

  \draw (0,0) ++(4,0)node(start)[draw,circle]{};
  \draw[-latex'] (start) ++(-1,0) -- (start);

  \draw (start) ++(4,0)node(1)[draw,circle]{};
  \draw[-latex'] (start) --node[above=-1]{
        \begin{tabular}{l@{\,\,\,\,}l}
        $-Max \ \overline{\mu}$	& $-Max \ \overline{i}$ \\
        $-Max \ \overline{\mu_L}$	& $-Max \ \overline{r}$ \\
        $-Max \ \overline{\mu_R}$	& $-Max \ \overline{t}$
      \end{tabular}} (1);

  \mybox{(1) ++(2,0)}{assign1}{
    \begin{minipage}{15mm}
      \hspace{0cm}
    \end{minipage}
  }{\emph{assign($\mu_L$, $l_v$)}}{1}

  \draw[-latex'] (1) -- (assign1);

  \mybox{(assign1) ++(4,0)}{assign2}{
    \begin{minipage}{15mm}
      \hspace{0cm}
    \end{minipage}
  }{\emph{assign($\mu$,$h_v$)}}{1}

  \draw[-latex'] (assign1) -- (assign2);

  \mybox{(assign2) ++ (4,0)}{assign3}{
    \begin{minipage}{15mm}
      \hspace{0cm}
    \end{minipage}
  }{\emph{assign($\mu_R$, $r_v$)}}{1}

  \draw[-latex'] (assign2) -- (assign3);

  \draw (assign3) ++(4,0)node(loop)[draw,circle]{};
  \draw[-latex'] (assign3) -- (loop);

\end{tikzpicture}

\caption{
Preliminary step of the reduction ($l_v$, $h_v$, and $r_v$ encode
the input tape configuration).}
\label{fig:preliminary_step}
\end{figure}

At this point, given a \lbatm $\mathcal A$ and an input tape configuration
$\atmtapeconf{s_0}$,
the game structure $G_{\mathcal A, \atmtapeconf{s_0}}$ presents, among others,
the following features
(the other features of $G_{\mathcal A, \atmtapeconf{s_0}}$ are either
irrelevant or represented in the graphical representation of the encoding
--- see \figurename{s}~\ref{fig:encoding_instr}-\ref{fig:preliminary_step}):
\begin{compactitem}
\item $2$ agents, $ag_1$ and $ag_2$;

 \item $5$ locations, namely $\fullstate{\atmstate q}{\mathsf{B}}$,
$\fullstate{\atmstate q}{\mathsf{1}}$, $\fullstate{\atmstate q}{\mathsf{2}}$,
$\fullstate{\atmstate q}{\llcorner}$, $\fullstate{\atmstate q}{\lrcorner}$,
for each internal state \atmstate q of $\mathcal A$
(plus other locations --- the circles in the pictures --- that do not correspond
to particular states of the $\lbatm$, but are needed to perform the encoding);

\item only one atomic proposition $p$, that holds true over all and only
the locations
having no matching instructions;

\item initial global availability $\vec{m_0}$ is such that
all resources are available in quantity $Max$, as already mentioned above;
notice that $Max$ also represents the maximum value occurring in the initial
resource availability vector, that is, $M = Max$;

\item initial location $\fullstate{\atmstate{q_0}}{\lambda}$,
where $\atmstate{q_0}$ is the initial state of the \lbatm and $\lambda$
is the first input symbol.

\end{compactitem}
The formula $\phi_{\mathcal A, \mathsf{s_0}} =
\team{A^{\vec{\money}}} \Diamond p$, with $A = \{ag_1\}$ and the value of
$\vec{\money}$ being irrelevant for our purposes,
is such that $G_{\mathcal A, \mathsf{s_0}} \models \phi_{\mathcal A, \mathsf{s_0}}$
if and only if $\mathcal A$ accepts on input $\atmtapeconf{s_0}$.

Notice that, for the sake of readability, the game structure used in the reduction does not
respect the requirement that, in every location, the first action of every agent is
the \emph{do-nothing} action, which does not consume or produce resources.
Nevertheless, this omission does not affect the correctness of our reduction, that can be easily
adapted using a game structure fulfilling the above requirement.
\ignoreconference{
 as follows.
We consider that all opponent's actions in the above game structure neither consume nor produce resources (so they
can be thought of as the \emph{do-nothing} action). Now, from the location where
proponent has no \emph{do-nothing} action (as all proponent's actions produce or consume
resources), we can introduce a \emph{do-nothing} action for the proponent, and
the resulting additional transition (corresponding to the \emph{do-nothing} action for both agents)
is defined as a loop transition, not consuming or producing resources and leading to the same location.
It is easy to see that $\phi_{\mathcal A, \mathsf{s_0}}$ is satisfied
in the original game structure if and only if it is satisfied
in the modified one.
}
\ignoreconference{
\begin{theorem}
The model checking problem for \prbatl is EXPTIME-hard even assuming $n$ and $r$ constant.
\end{theorem}
}
\begin{theorem}
Model checking \prbatl is EXPTIME-hard even assuming $n$ and $r$ constant.
\end{theorem}

Let us stress that the above reduction makes use of a constant
number of agents and resources, while the digit size of
$M$ (the maximum value occurring in 
$\vec{m_0}$) is linear in the size of the
tape configuration.
This is consistent with the complexity of the algorithm in \cite{dnp11}, which
remains exponential even if we consider a constant number of agents and resources as input.

\begin{corollary}
The model checking problem for \prbatl is EXPTIME-complete.
\end{corollary}

\subsection{Another reduction.}\label{constAgAndM}
As noted at the beginning of Section~\ref{hardnessMC}, it is possible to exhibit two more 
reductions
according to which two parameters, out of three, are assumed constant.
%
In the following, we briefly outline how to obtain a reduction from the acceptance 
problem for \lbatm, when $n$ and $M$ are constant. 

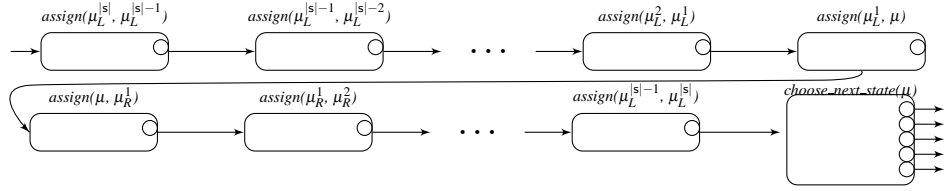
\begin{figure}[t]
\tiny
\centering
\begin{tikzpicture}[scale=.5]
\draw (-1,1.2) ++(25,-4.5);

  \mybox{(0,0)}{assign}{
    \begin{minipage}{15mm}
      \hspace{0cm}
    \end{minipage}
  }{\emph{assign($\mu_L^{|\atmstate s|}$, $\mu_L^{|\atmstate s|-1}$)}}{1}

  \draw[-latex'] (assign) ++(-2.5,0) -- (assign);

  \mybox{(assign) ++(4,0)}{assign2}{
    \begin{minipage}{15mm}
      \hspace{0cm}
    \end{minipage}
  }{\emph{assign($\mu_L^{|\atmstate s|-1}$, $\mu_L^{|\atmstate s|-2}$)}}{1}

  \draw[-latex'] (assign) -- (assign2);

  \draw (assign2) ++(4.5,0)node[inner sep=10](dots){\Large $\ldots$};
  \draw[-latex'] (assign2) -- (dots);

  \mybox{(dots) ++(2.5,0)}{assign3}{
    \begin{minipage}{15mm}
      \hspace{0cm}
    \end{minipage}
  }{\emph{assign($\mu_L^{2}$, $\mu_L^{1}$)}}{1}

  \draw[-latex'] (dots) -- (assign3);

  \mybox{(assign3) ++(4,0)}{assign4}{
    \begin{minipage}{15mm}
      \hspace{0cm}
    \end{minipage}
  }{\emph{assign($\mu_L^{1}$, $\mu$)}}{1}

  \draw[-latex'] (assign3) -- (assign4);

  \mybox{(assign) ++(-2,-2.15)}{assign5}{
    \begin{minipage}{15mm}
      \hspace{0cm}
    \end{minipage}
  }{\emph{assign($\mu$, $\mu_R^1$)}}{1}

  \draw (assign4.south) ++(-1,-.25)node(c1){} (assign5.west) ++(0,1.25)node(c2){};
  \draw plot [cyan,smooth, tension=0.2] coordinates {
  (assign4.south) (c1) (c2) (assign5.west) };
  \draw[-latex'] (assign5.west) ++ (-.1,.1) -- (assign5.west);

  \mybox{(assign5) ++(4,0)}{assign6}{
    \begin{minipage}{15mm}
      \hspace{0cm}
    \end{minipage}
  }{\emph{assign($\mu_R^{1}$, $\mu_R^{2}$)}}{1}

  \draw[-latex'] (assign5) -- (assign6);

  \draw (assign6) ++(4.5,0)node[inner sep=10](dots2){\Large $\ldots$};

  \draw[-latex'] (assign6) -- (dots2);

  \mybox{(dots2) ++(2.5,0)}{assign7}{
    \begin{minipage}{15mm}
      \hspace{0cm}
    \end{minipage}
  }{\emph{assign($\mu_L^{|\atmstate s|-1}$, $\mu_L^{|\atmstate s|}$)}}{1}

  \draw[-latex'] (dots2) -- (assign7);


  \draw (assign7) ++(4,0)node(chooseTemp){};
  \myboxCustom{(assign7) ++(4,-.2)}{choose}{
    \begin{minipage}{15mm}
      \hspace{0cm}
    \end{minipage}
  }{\emph{choose\_next\_state($\mu$)}}{5}{12mm}

  \draw[-latex'] (assign7) -- (chooseTemp);

  \draw[-latex'] (choose_out1) -- ++(1,0);
  \draw[-latex'] (choose_out2) -- ++(1,0);
  \draw[-latex'] (choose_out3) -- ++(1,0);
  \draw[-latex'] (choose_out4) -- ++(1,0);
  \draw[-latex'] (choose_out5) -- ++(1,0);

\end{tikzpicture}

\caption{Encoding of the module \emph{shift\_right}.}
\label{fig:module_shift_right3}
\end{figure}

This reduction is simpler than the previous. Here the encoding of the tape
is obtained using a number of resources which is linear in the length of the tape.
Let $| \mathsf{s} |$ be the length of the tape, we use $2$ sets of
$|\mathsf{s}|$ resource variables, namely, $\mu_L^1, \mu_L^2, \ldots, \mu_L^{|\mathsf{s}|}$
and $\mu_R^1, \mu_R^2, \ldots, \mu_R^{|\mathsf{s}|}$, plus the resource variable $\mu$.
Each variable encodes the content of a tape cell: variable $\mu$ encodes the
content of the head cell, while, for each $i$, the variable $\mu_L^i$ (resp.,
$\mu_R^i$) encodes the content of the $i$-th cell on the left (resp., right)
of the tape cell.
Notice that, since there are finitely many possible values for a tape cell,
the value of $M$ is upper bounded.
Now, the encoding of the set of instructions matching a full state
\fullstate{\atmstate q}{\lambda} of a \lbatm is the same used for the previous
reduction and depicted in \figurename~\ref{fig:encoding_instr}.
Nevertheless, the encoding of the module \emph{move},
which encodes the shift (to right or to left) of the head, is slightly different.
In \figurename~\ref{fig:module_shift_right3}, the sub-modules encoding the operation
``shift to right'' are depicted.
Essentially, the value of the variable representing a cell is transmitted to the variable
representing the cell on the right, and the next location reached on the game structure
is set according to the value stored on the current head cell (after the shift operation).
The encoding of the operation ``shift to left'' is made analogously.

\ignoreconference{
\begin{theorem}
The model checking problem for \prbatl is EXPTIME-hard even assuming $M$ and $r$ constant.
\end{theorem}
}
\begin{theorem}
Model checking \prbatl is EXPTIME-hard even assuming $n$ and $M$ constant.
\end{theorem}


\section{Discussion}
In this paper we have presented a formalism which is very suitable to model properties of
multi-agent systems when the agents share resources  and the need of avoiding an unbounded consumption of 
such resources is crucial. Within our framework it is possible to keep trace of a real global availability
of the resources, used by both the proponent and opponent players, avoiding thus unrealistic
situations in which an unbounded quantity of resources is used in a game.
 
 The technical focus of the paper  has been on the complexity of the model checking problem, and we proved that  it is 
EXPTIME complete (recall that also for simpler  formalism 
this problem is in EXPTIME, though the lower bound is not known). 
Other  problems of interest exist in the context of multi-agents system verification.
The most important one is the \emph{reachability problem}, that is 
the problem of determining whether a team,  with a given amount of money and a given
 initial global resource  availability,  has a strategy to force the execution of the system to reach a given location.
More precisely,  the reachability problem for a team $A$ on a priced game structure
$G$ is a particular instance of the model checking problem, namely,
the problem of verifying the truth at the initial configuration of $G$
of a \prbatl\ formula of the kind $\team{A^{\vec \$}} \Diamond p$, for a team $A$, 
a money endowment ${\vec \$}$ and  $p \in \Pi$.
An upper bound on the complexity of this problem is clearly given by the algorithm for solving 
the model checking problem for  \prbatl.
Let us observe that the reductions given in section~\ref{hardnessMC} apply also to the reachability problem, 
since the formula used there was $\phi_{\mathcal A, \mathsf{s_0}} = \team{A^{\vec{\money}}} \Diamond p$,
thus we have the following corollary.
\begin{corollary}
The reachability problem for \prbatl is EXPTIME-complete.
\end{corollary}

One of the novelties of our logic is that {\em the resource production} is allowed in the actions, 
though with some limitations. 
Model checking and reachability problems seem both to be simpler in the case one restricts our formalism 
by considering  agent actions that cannot produce resources.
The reachability problem is indeed NP-hard in this case: 
it immediately  follows from a result in \cite{DBLP:conf/ceemas/JamrogaD05}, when the number of agents 
is not constant. Anyway, we can prove the NP-hardness for just two agents using a reduction from 3-SAT
(due to lack of space we omit here the proof).
The model checking problem, instead, turns out to be PSPACE-hard, since 
the reduction from  QBF problem 
given in \cite{dnp11} works also in this case, when actions cannot  produce resources. 
Observe that  \prbatl\  with this restriction  is again  different from other formalisms in literature, 
mainly for the possibility of tracking resources avalability and for considering shared resources. 

Finally, we want to note that also the more general problem, called  \emph{optimal coalition problem}, 
is EXPTIME-complete (the upper bound was shown in \cite{dnp11}).
It is the problem of finding optimal (with respect to a suitable cost function) coalitions
that are capable to satisfy a given  \emph{parametric \prbatl} formula,
that is, a \prbatl formula in which \emph{parametric team operators}
$\team{X^{\vec{\money}}}$ may occur in place of the classical team
operators $\team{A^{\vec{\money}}}$. 
One could also investigate other optimization problems. The most interesting is, perhaps,
to  consider the money
availability not as an input of the problem, but rather as a parameter to minimize, that is
to establish how much money each agent should be provided with, to perform a given task.

Further research directions concern the study of variants of the logic. First, one can consider 
extensions based on the full alternating-time temporal
language \atlstar,  as already done in \cite{BF10}, and  its fragment \atlplus.

%
%

\bibliographystyle{eptcs}
\bibliography{biblio}

\end{document}